%% file: main.tex
\documentclass[aps, prb, floatfix, twocolumn, showpacs, superscriptaddress, longbibliography]{revtex4-1} 

\usepackage{amsmath, amsfonts, amssymb, dsfont}
\usepackage{mathtools}
\usepackage[smalltableaux]{ytableau}
\usepackage{graphicx}
\usepackage[export]{adjustbox}
\usepackage{array}
\usepackage[toc,page,titletoc]{appendix}
\usepackage{pgfplots}
\usepackage{standalone}
\usepackage{import}
\usepackage[hypertexnames=false]{hyperref}
\usepackage[capitalize]{cleveref} 
\usepackage[normalem]{ulem}
\usepackage{subfiles}
\usepackage{physics}
\usepgfplotslibrary{groupplots}
\pgfplotsset{compat=newest}
\hypersetup{
   colorlinks,
   citecolor=blue,
   filecolor=blue,
   linkcolor=blue,
   urlcolor=blue,
   breaklinks=true
}
\usepackage{color}  

\usepackage{tikz}
\usetikzlibrary{matrix,backgrounds}
\newcommand*\expect[1]{\left\langle #1  \right\rangle}

\input{tikzfig}

\begin{document}
\title{Singularity with and without disorder at AKLT points}
\author{Loïc Herviou}
\affiliation{Univ. Grenoble Alpes, CNRS, LPMMC, 38000 Grenoble, France}
\author{Anthony Rey}
\affiliation{Laboratory for Theoretical and Computational Physics, PSI Center for Scientific Computing, Theory and Data, Paul Scherrer Institute, 5232 Villigen,
Switzerland}
\author{Fr\'ed\'eric Mila}
\affiliation{Institute of Physics,  École Polytechnique Fédérale de Lausanne (EPFL), CH-105 Lausanne, Switzerland}

\begin{abstract}
The Affleck-Kennedy-Lieb-Tasaki (AKLT) point of the bilinear-biquadratic spin-1 chain is a cornerstone example of a disorder point where short-range correlations become incommensurate, and correlation lengths and momenta are non-analytic. While the presence of singularities appears to be generic for AKLT points, we show that for a family of SU(N) models, the AKLT point is not a disorder point: It occurs entirely within an incommensurate phase yet the wave vector remains singular on both sides of the AKLT point. We conjecture that this new possibility is generic for models where the representation is not self-conjugate and the transfer matrix non-Hermitian, while for self-conjugate representations the AKLT points remain disorder points.
\end{abstract}

\maketitle

\paragraph*{Introduction.}
The discovery in 1987 by Affleck, Lieb, Kennedy and Tasaki of a special spin-1 chain with biquadratic interactions\cite{AKLT} known as the AKTL point has played a very important role in condensed matter physics. 
It was the first example of a spin-1 chain for which the presence of a gap predicted by Haldane\cite{HaldaneConjecture1,HaldaneConjecture2} could be rigorously proven analytically\cite{AKLT2}, and the simple valence-bond solid structure of the ground state at that point has paved the way to the modern formulation of quantum models in terms of Matrix Product States (MPS)\cite{SchollwockDMRG} or of topological phases\citep{Lauchli2006, DeChiara2011, Kennedy1992, Pollmann2010}. 
It has also impacted theoretical condensed matter physics in higher dimension through the successful generalisation of valence-bond solid states to 2D models\cite{book_auerbach} and of the MPS to tensor networks\cite{verstraete1,verstraete2}. 

Interestingly enough, a connection between the AKLT model and commensurate-incommensurate transitions has also been established in the case of the spin-1 chain\cite{Schollwock1996}. 
Upon increasing the biquadratic interaction, the AKLT point appears as a disorder point where short-range correlations become incommensurate, with the typical properties of such a point known from 2D classical physics\cite{DenNijs}: the correlation length has a kink and the wave-vector grows as a square-root from the commensurate phase. 
The same properties have been found at the Majumdar-Ghosh point\cite{Majumdar1969} of the $ J_1-J_2$ spin-1/2 chain\cite{White1996}, for which the ground state also has a very simple structure, a product state. 
It is therefore tempting to conjecture that these properties will be realized at all the AKLT points that can be constructed as generalisations of the AKLT point to larger spins or to other symmetry groups such as SU$(n)$.

In this Letter, we show that this conjecture is not completely true. 
We have investigated the AKLT point of several SU$(n)$ models, and we have found that, while they all exhibit a kink of the correlation length and a singularity of the wave-vector of the short-range correlations, they do not all correspond to a disorder point. 
For instance, the short-range correlations of the 3-box symmetric SU(3) model are incommensurate on both sides of the AKLT point, and the singularity appears as an infinite slope of the wave-vector at the AKLT point, with square-root singularities on both sides. 
Our new conjecture is that AKLT points are {\it bona fide} disorder points if the transfer matrix is Hermitian, which is the case if the representation is self-conjugate, as for the spin-1 SU(2) model, and that they are singular points between phases with incommensurate correlations if the transfer matrix is non-Hermitian, which is the case if the representation is not self-conjugate, as for the 3-box symmetric representation of SU(3).
The discovery of a singularity within an incommensurate gapped phase introduces a novel and unexpected feature in the landscape of quantum spin chains and of two-dimensional classical models.\\

\paragraph*{Generalized SU$(n)$ AKLT chains.}
We consider SU$(n)$ spins chains. 
For simplicity, we consider the case of a single irreducible representation (irrep), denoted $\vb*{p}$, on each physical site (see App.~\ref{app:basics} for a brief reminder on the irreducible representations of SU$(n)$).
An AKLT state is a nearest-neighbour valence bound state with singlets forming on each bond.
Let $\vb*{v_L}$ (resp. $ \vb*{v_R})$ be the irrep forming the singlet on the left (resp. right) bond.
In the thermodynamic limit, this state is an exact matrix product state:
\begin{equation}
    \ket{\vb*{p} ; \vb*{v_L}, \vb*{v_R}} \approx ... M^\sigma_1 M^\sigma_2 M^\sigma_3 ...
\end{equation}
where $M^\sigma$ is the $3$-tensor
\begin{equation}
    \localtensor
\end{equation}
chosen such that $\ket{\vb*{p} ; \vb*{v_L}, \vb*{v_R}}$ is normalized.
To be a valid state, the trivial irrep $\bullet$ and $\vb*{p}$ must appear in the decomposition of $\vb*{v_L} \otimes \vb*{v_R}$.

Given such a state, we can derive an exact parent Hamiltonian in the following way\citep{Gozel2019}.
We restrict ourselves here to $2$-site Hamiltonians.
We decompose the pair of irreps into
\begin{equation}
    \vb*{p} \otimes \vb*{p} = \bigoplus\limits_j n_j \vb*{p_j}, \quad \vb*{v_L} \otimes \vb*{v_R} = \bigoplus\limits_k m_k \vb*{v_k}.
\end{equation}
Let $\mathcal{I}$ be the intersection of $\{\vb*{p_j}\}$ and $\{\vb*{v_k}\}$ and $\overline{\mathcal{I}}$ the remainder of $\{\vb*{p_j}\}$.
We are looking for a local operator with finite positive weights on $\overline{\mathcal{I}}$ and zero weights in $\mathcal{I}$.
We denote $\vec{T}_j$ the generators of $\mathfrak{su}(n)$ for the irrep $\vb*{p}$ at site $j$ and $\vec{T}_{j + \frac{1}{2}} = \vec{T}_{j} + \vec{T}_{j+1} $ the total spin operators for two sites.
$C(\vb*{i})$ is the quadratic Casimir operator for the irrep $\vb*{i}$.
The Hermitian operator 
\begin{equation}
    h_{j+\frac{1}{2}} = \prod\limits_{C(\vb*{i}) \in C(\mathcal{I})} \left( \vec{T}_{j+\frac{1}{2}}^2 - C(\vb*{i}) \right)
\end{equation}
has weight $0$ on all irreps in $\mathcal{I}$.
In the models considered in this paper, it is positive semi-definite and $\ket{\vb*{p} ; \vb*{v_L}, \vb*{v_R}}$ is an exact zero-energy state of $H = 
\sum_j
h_{j+\frac{1}{2}}$ \footnote{We can always choose  $H_2 = \sum\limits_j h_{j+\frac{1}{2}}^2$ if $h_{j+\frac{1}{2}} $ is not positive definite. }, and up to edge states, the ground state is unique.
Relevant examples for this construction are given in App.~\ref{app:parent}.\\

\paragraph*{Methods: matrix product states and C-IC correlation.}
We use infinite MPS (iMPS)\citep{SchollwockDMRG} to represent the different groundstates. 
As a brief reminder, in its mixed canonical norm, an iMPS is a set of tensors such that the wavefunction in the thermodynamic limit can be written as
\begin{equation}
    \ket{\psi} \approx ... L_1...L_N L_1...L_N C R_1...R_N R_1...R_N...,
\end{equation}
where $N$ is the size of the unit-cell ($N = 2$ in the following), $C$ a center matrix and $L_j$ (resp. $R_j$) are left- (resp. right-) normalized $3$-tensors. 
Computations are performed using the open-source library \href{https://github.com/Jutho/TensorKit.jl}{TensorKit.jl}\citep{TensorKit} and \href{https://github.com/QuantumKitHub/MPSKit.jl}{MPSKit.jl}\citep{MPSKIT}. 
The SU$(n)$ symmetry is implemented using \href{https://github.com/QuantumKitHub/SUNRepresentations.jl}{SUNRepresentations.jl} based on Ref.~\onlinecite{Alex2011}.
We obtain the groundstates using a combination of iDMRG\citep{WhiteDMRG, VidaliDMRG, McCullochiDMRG}, VUMPS\citep{VUMPS} and Grassmann gradient descent\citep{GrassmannDescent}.
At a given bond dimension, we first perform two-site iDMRG (only for small bond dimensions).
The initialization is performed with boundary conditions such that the edge states are screened.
We then switch to VUMPS, followed by steps of Grassmann gradient descent to converge to high accuracy.
Our final tolerance is of order $10^{-10}$ on the energy, though we truncate singular values below $10^{-12}$.
We use optimal subspace extension\citep{SubspaceExpansion} to enlarge the bond dimension.

For the gapped one-dimensional models we study, the generic two-point connected correlation function for arbitrary local operators $\hat{O}_x$ defined at site $x$ takes the Ornstein-Zernike\citep{Ornstein1914} form
\begin{equation}
    \expect{\hat{O}_x \hat{O}'_y}_c \approx \frac{e^{-\frac{y-x}{\xi_{\hat{O} \hat{O}'}}}}{\vert y-x \vert^{\eta_{\hat{O} \hat{O}'}} } \cos q_{\hat{O} \hat{O}'} (y-x + \phi_{\hat{O} \hat{O}'}),\label{OZ}
\end{equation}
where $\xi_{\hat{O} \hat{O}'}$ is the associated correlation length, $q_{\hat{O} \hat{O}'}$ the corresponding wavevector and $\eta_{\hat{O} \hat{O}'} \geq 0$ a non-universal exponent.
If $q = \frac{2\pi m}{N}$ with $N$ the unit-cell size and $m \in \mathbb{Z}$, the correlations are commensurate.
Otherwise, they are incommensurate.

The iMPS representation of a state is a convenient tool to characterize the long-range behavior of such correlators.
We define the left transfer matrix $T_L$ as 
\begin{equation}
    T_L = \transfermatrix 
\end{equation}
It is a non-Hermitian matrix of dimension $\chi^2 \times \chi^2$, where $\chi$ is the bond dimension of the iMPS.
It is always time-reversal symmetric such that
\begin{equation}
    \sigma^x T_L^* \sigma^x = T_L,
\end{equation}
where $\sigma^x$ exchange the top and bottom legs, and its eigenvalues come in conjugate pairs.
Any generalized two-point correlation function can be readily evaluated using
\begin{equation}
    \expect{\vec{O}_x \cdot \vec{O}_y} = \sum\limits_\alpha \expect{\hat{O}^\alpha_x  \hat{O}^\alpha_y}  = \left( \vec{O}_x \vert T_L^{\frac{y - x}{2}} \vert \vec{O}_y \right)
\end{equation}
\[= \leftvector  \left( \bigtransfermatrix \right)^{\frac{y - x}{2}}  \rightvector\]
Here $x$ and $y$ are physical positions and $\alpha$ labels different local operators.
The right leg of $\vec{O}_x$ is contracted with the left leg of $\vec{O}_y$ to represent the sum over $\alpha$.
Note that in this example, we assumed $x$ was even and $y$ odd, but it generalizes trivially.
We can therefore diagonalize the transfer matrix into
\begin{equation}
    T_L = \left. \vert R_0 \right) \left( L_0\vert \right. + \sum\limits_n t_n \left. \vert R_n \right) \left( L_n\vert \right., \, 1 \geq \vert t_1 \vert \geq  \vert t_2 \vert...
\end{equation}
$\left. \vert R_n \right)$ (resp. $\left( L_n\vert \right.$) are the right- (resp. left-) eigenvectors of $T_L$.
Eigenvalues of norm $1$ contribute to the disconnected part of the correlator, or to a non-trivial order parameter, while the exponential decay arises from $\vert t_n \vert < 1$.
In the states we consider, $\vert t_1\vert < 1$.
The operator-dependent correlation lengths and momenta are therefore simply given by
\begin{equation}
\xi_{\vec{O} \cdot \vec{O}}^{(n)} = -\frac{N}{\log \vert t_n \vert},~ q_{\vec{O} \cdot \vec{O}}^{(n)} = \frac{\text{Im}\log t_n}{N}\,\text{ mod } \frac{2\pi}{N}. 
\end{equation}
under the constraint that the structure factor
\begin{equation}
        w_{\vec{O} \cdot \vec{O}}^{(n)} = \left( \vec{O}_x \vert  R_n \right) \left( L_n \vert \vec{O}_y \right) \neq 0.
\end{equation}
The correlations of MPS with finite bond dimension are always strictly exponential.
An Ornstein-Zernike form with $\eta > 0$ is only recovered in the limit of infinite bond dimension as it decomposes into an (infinite) sum of exponentials\citep{Zauner2015, Rams2018}.
This subtlety is largely irrelevant close to the AKLT point where $\eta = 0$, and does not affect our analysis.
Due to the small correlation lengths in our model, we will not perform the standard extrapolation schemes\citep{Rams2018, Vanhecke2019} as finite-bond effects are dominated by the numerical instability and noise in the computation of the transfer matrix.

The irreps on the horizontal legs of $\vec{O}_{x/y}$ determine which sectors of $T_L$ to study.
If $N > 1$ and the spectrum of the transfer matrix is not degenerate, it can be convenient to use
\begin{equation}
 \tilde{q}_{\vec{O} \cdot \vec{O}}^{(n)} =\text{Im}\log  \frac{\left( \vec{O}_x \vert  R_n \right) \left( L_n \vert \vec{O}_y \right)}{\left( \vec{O}_{x+1} \vert  R_n \right) \left( L_n \vert \vec{O}_y \right)}. \label{eq:ratio}
\end{equation}
Indeed, in an ideal system, $q$ and $\tilde{q}$ should be equal if translation invariance is not broken.
In practice we use $\tilde{q}$ only to determine the value of $q$ modulo $2\pi$.\\

As a concluding remark, we point out that the transfer matrix and its eigenvalues are generically numerically unstable, due to its non-Hermitian nature.
Small errors on the local tensors can be amplified, in particular for larger bond dimensions or close to exceptional points.
A benchmark of our method for the SU$(2)$ spin-1 AKLT chain is shown in App.~\ref{app:SU2}.\\

\paragraph*{SU$(n)$ generalizations with self-conjugate representations.}
The first direct generalization of the spin-1 AKLT chain corresponds to the states $\ket{\vb*{n^2-1}; \vb*{n}, \vb*{\overline{n}}}$\citep{Greiter2007,Greiter2007b,Katsura2008}.
A local on-site SU$(n)$ adjoint irrep is decomposed into a SU$(n)$ fundamental representation, $\vb*{n}$, and its conjugate, $\vb*{\overline{n}}$, on the bonds.
The corresponding parent Hamiltonian is 
\begin{equation}
    H_{\mathrm{AKLT}} = \sum\limits \vec{T}_j \cdot \vec{T}_{j+1} + \beta \sum\limits (\vec{T}_j \cdot \vec{T}_{j+1})^2, \label{eq:parentSUN}
\end{equation}
with $\vec{T}$ the generators of SU$(n)$ in the defining representation and $\beta_\mathrm{AKLT}  = \frac{2}{3n}$ (see App.~\ref{app:parent}).
With open boundaries, there are $n^2$ topologically stable edge states corresponding to the representation $\vb*{n} \otimes \vb*{\overline{n}}$\citep{Morimoto2014}.

The entanglement spectrum at the AKLT point in the bulk is a single $n$-uplet of weight $n^{-\frac{1}{2}}$.
The corresponding two-site transfer matrix has only two distinct eigenvalues: $1$ in the trivial representation $\bullet$ and $\frac{1}{(n^2-1)^2}$ in the adjoint irrep.
The latter controls the spin-spin correlator such that 
\begin{equation}
 \expect{\vec{T}_x \cdot \vec{T}_y}_c \propto \left(\frac{-1}{n^2-1} \right)^{\vert x-y \vert},
\end{equation}
i.e. the correlations are commensurate with momentum $q_{\vec{T}\cdot\vec{T}}^{(1)} = \pi $ and correlation length $\xi_{\vec{T}\cdot\vec{T}}^{(1)} = \frac{1}{\log (n^2-1 )}$.

We vary $\beta$ around the AKLT point for $n =3$ and $n = 4$.
Remarkably, even close to the AKLT point, the dominant eigenvalues of the transfer matrix in the adjoint irrep are significantly larger than at the AKLT point (see Fig.~\ref{fig:SU3-all} in Appendix).
In fact, the transfer matrix spectrum looks discontinuous at the AKLT point both at small and large bond dimensions, even though the correlation functions are supposed to evolve continuously.
This apparent contradiction is resolved by a careful study of the correlation functions and the transfer matrix.
The weights $w_{\vec{T}. \vec{T}}^{(n)}$ go to $0$ for the $\vert t_n \vert> t_\mathrm{AKLT} $ that do not converge towards $t_\mathrm{AKLT} $.
These largest eigenvalues dominantly contribute to the correlations when the amplitude of the connected correlator reaches approximately $10^{-8}$ or below.
They therefore describe a non-physical regime caused by the finite accuracy of the iMPS and the finite precision of the floating-point algebra.
Note that, paradoxically, this effect is visible due to the very short correlation lengths and the simplicity of the wavefunctions close to the AKLT points.

\begin{figure}
    \centering
    \includegraphics[width=\linewidth]{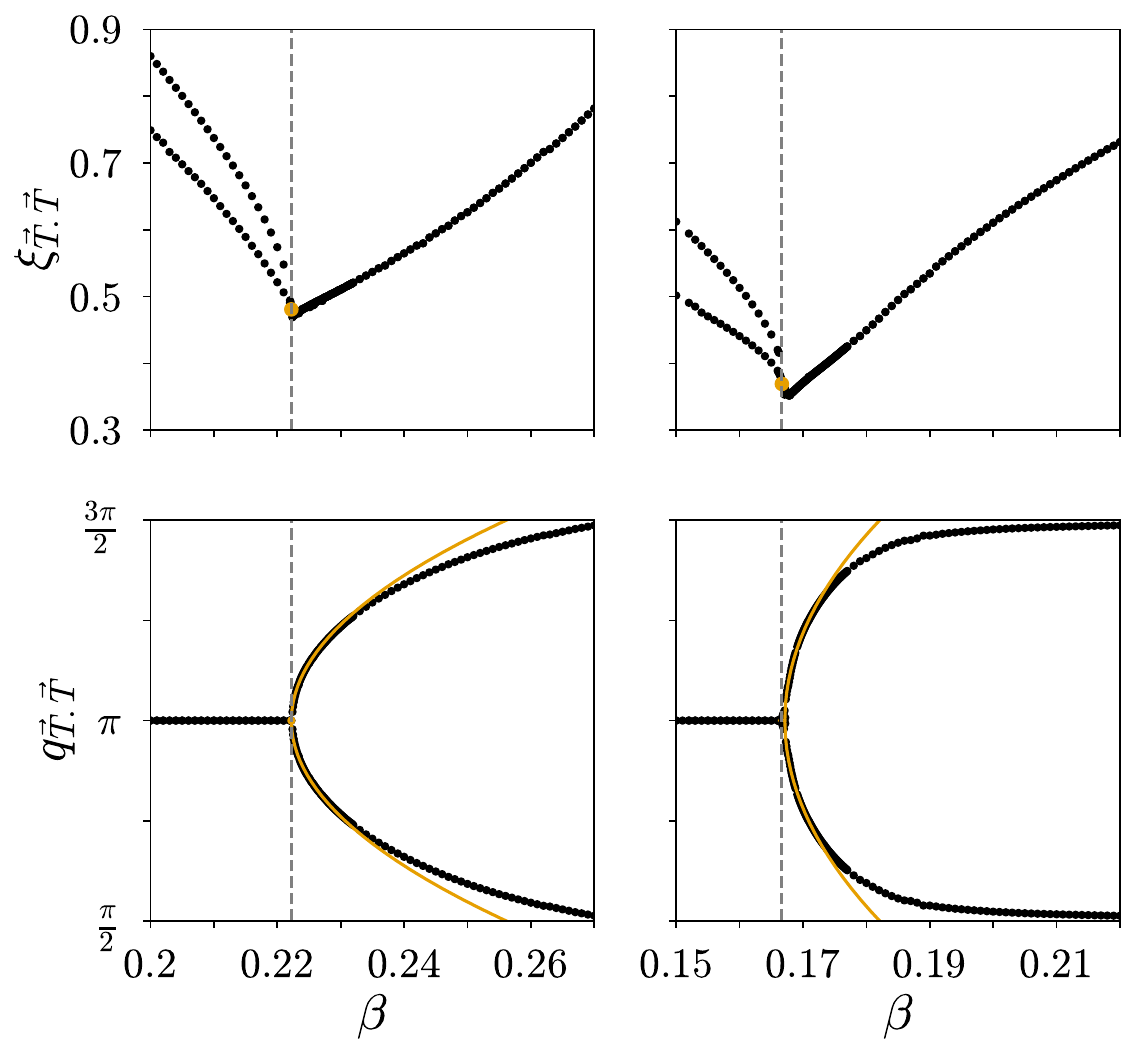}
    \caption{Top: the two largest correlation lengths associated to the spin-spin correlations following Eq.~\eqref{eq:recipe} for SU$(3)$ (left) and SU$(4)$ (right). The dashed grey line marks the AKLT point. In both cases, the two eigenvalues converge towards their AKLT value, despite the existence of a single eigenvalue at the disorder point. Note that the small horizontal shift of the kink for SU$(4)$ is of order $5\times 10^{-4}$ and is likely an artefact of our finite precision. Bottom: momentum associated to these two eigenvalues. The square root behavior (orange fit) is typical of the coalescence of eigenvalues in non-Hermitian matrices.}
    \label{fig:Zn}
\end{figure}

Several selection schemes can be used to select the physical eigenvalues of the transfer matrix.
As weight $ w_{\hat{O} \hat{O}'}^{(n)}$ and amplitude $\vert  t_{\hat{O} \hat{O}'}^{(n)} \vert $ are relevant parameters, we associate to each eigenvalue the distance $d_{\hat{O} \hat{O}'}^{(n)}$ that verifies
\[
\vert  w_{\hat{O} \hat{O}'}^{(n)}  \left(  t_{\hat{O} \hat{O}'}^{(n)} \right)^{d_{\hat{O} \hat{O}'}^{(n)}} \vert = \lambda_\mathrm{th}, \label{eq:recipe}
\]
where the threshold $\lambda_\mathrm{th} = 10^{-5}$ is a free parameter.
$d_{\hat{O} \hat{O}'}^{(n)}$ corresponds to the distance at which the contribution of the eigenvalue to the connected correlations reaches $\lambda_\mathrm{th}$.
Eigenvalues with the largest $d_{\hat{O} \hat{O}'}^{(n)}$ while $\lambda_\mathrm{th}$ remains above the numerical noise control the long-range behavior of the correlations.
%
We verified that the dominant eigenvalues are essentially independent of $\lambda_\mathrm{th}$ in the range $10^{-6} \leq \lambda_\mathrm{th} \leq 10^{-4}$.
Nonetheless, they can be relatively deep in the spectrum of the transfer matrix, hence a reduced accuracy due to numerical instability.
In Fig.~\ref{fig:Zn}, we represent the correlation lengths and the momenta associated to the two eigenvalues that maximize $d_{\vec{T}\cdot\vec{T}}$.
They appear to coalesce at the AKLT point, such that the correlation length evolves continuously with a kink at the transition.
The momentum is commensurate ($\pi$) for $\beta \leq \beta_\mathrm{AKLT}$, but becomes incommensurate for $\beta > \beta_\mathrm{AKLT}$, with the typical square-root behavior $q_{\vec{T}\cdot\vec{T}} - q_{\mathrm{AKLT}} \propto \sqrt{\beta -  \beta_\mathrm{AKLT}}$. 
Consequently, this family of AKLT points are disorder points marking a commensurate-incommensurate transition.
We have obtained similar results for SU$(5)$ (see  App.~\ref{app:AddData}), but they could not be converged to the same precision due to the limitation in getting Clebsch-Gordan coefficients for very large irreps.\\

\paragraph*{AKLT with non self-conjugate representations.}
Instead of having a fundamental representation and its adjoint forming singlets on a bond, it is possible to look for AKLT states with virtual adjoint irreps.
In the SU$(3)$ case, this construction corresponds to the $3$-box  symmetric AKLT state studied in Refs~\onlinecite{Greiter2007, Gozel2019, Gozel2020, Devos2022}.
The physical Hilbert space is the 10-dimensional symmetric irrep $\vb*{10} = \ydiagram{3}$.
The parent Hamiltonian is also given by Eq.~\eqref{eq:parentSUN}, with $\beta_\mathrm{AKLT}  = \frac{1}{5}$ (see App.~\ref{app:parent}).
The irrep controlling the spin-spin correlator is also the SU(3) adjoint $\vb*{8}$.

In previous models, the transfer matrix at the AKLT point was Hermitian because the physical irrep $\vb*p$ was self-conjugate (proof in App.~\ref{app:TM}).
It is no longer the case for $\vb*{10}$.
The left legs of the transfer matrix decompose as
\begin{equation}
    \vb*{8}\otimes \vb*{8} = \bullet \oplus 2\times\vb*{8} \oplus \vb*{10} \oplus \vb*{\overline{10}} \oplus  \left( \vb*{27} = \ydiagram{4, 2} \right).
\end{equation} 
The eigenvalues of the single-site transfer matrix in the adjoint irrep are the complex conjugate pair $-0.2 \pm 0.4i$, with $q_{\vec{T}\cdot\vec{T}} = \pm ( \pi - \arctan 2) $ and $\xi_{\vec{T}\cdot\vec{T}} = \frac{2}{\log 5}$.
The correlations are already incommensurate at the AKLT point, so a continuous commensurate-incommensurate transition cannot occur at the AKLT point in the $3$-box symmetric model.

\begin{figure}
    \begin{center}
    \includegraphics[width=0.95\linewidth]{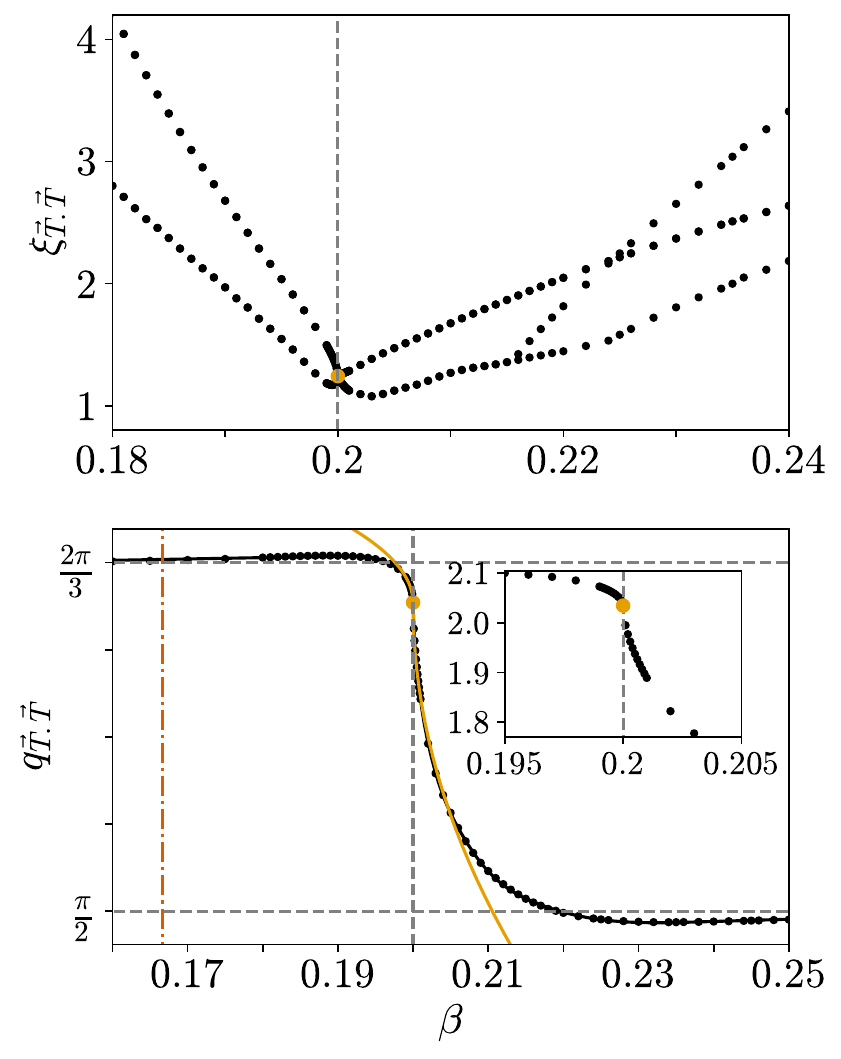}
    \end{center}
    \caption{Top: correlation lengths of the four largest eigenvalues of the transfer matrix for the $3$-box symmetric model. Bottom: momenta associated to these eigenvalues, with square root fit in orange. Two conjugate pairs of eigenvalues coalesce at the AKLT point, which remains a disorder point. It now separates two incommensurate regimes. Note that the additional crossovers withing the second incommensurate phase. In inset, a zoom to the vicinity of the AKLT point. The red vertical line is the classical onset of incommensuration: the chain is already incommensurable there.}
    \label{fig:Adjoint}
\end{figure}

Nonetheless, the AKLT point is non-trivial.
In Fig.~\ref{fig:Adjoint}, we represent the four eigenvalues of the transfer matrix with largest amplitude.
Unlike in the self-conjugate case, the dominant eigenvalues are physical and have large non-zero weights.
At the AKLT point, we see that two pairs of conjugate eigenvalues coalesce, leading to a kink in the correlation length.
Remarkably, while they are already incommensurate at the AKLT point, we do observe an asymmetric divergence of the derivative of $q_{\vec{T}\cdot\vec{T}}$.
Numerical fits give $q_{\vec{T}\cdot\vec{T}} - q_\mathrm{AKLT} \equiv \alpha \sqrt{\vert \beta_\mathrm{AKLT} - \beta \vert } $, with $\alpha \approx 1.2$ if $\beta < \beta_\mathrm{AKLT}$ and $-4.2$ if  $\beta > \beta_\mathrm{AKLT}$.
The AKLT point in this model has all the hallmark features of a disorder point (kink in the correlation length and singularity of the momenta), but does not separate a commensurate and an incommensurate regime.
To our knowledge, it is the first time that such a singularity is observed.

Semi-classical and flavour-wave approaches of the Heisenberg SU$(3)$ 3-box symmetric models have been proposed\citep{Lajko2017, Wamer2020}.
The semi-classical approach can be readily generalized to our model (see App.~\ref{app:classical}).
It suggests the existence of a commensurate-incommensurate transition, as the minimizing classical field acquires a finite momentum at a critical $\beta_\mathrm{class} < \beta_\mathrm{AKLT}$, though, it does not predict the incommensurate-incommensurate singularity.
Our simulations indeed indicate that the chain is incommensurate at $\beta_\mathrm{class}$, and that the commensurate wave-vector $q_{\vec{T}\cdot\vec{T}} = 2 \pi/3$ is only reached at $\beta = 0$.

We conjecture that this situation is generic when the transfer matrix is not Hermitian at the AKLT point and the AKLT point is therefore already incommensurate, such as the SU($4$) state $\vert \vb*{45}; \vb*{15}, \vb*{15} \rangle $.
Note that the presence of two copies of the adjoint in the transfer matrix does not guarantee incommensuration.
The SU$(3)$ AKLT state $\vert  \vb*{27}; \vb*{8}, \vb*{8} \rangle$, for example, is commensurate as $\vb*{27}$ is self-conjugate, and preliminary numerics indicate a conventional C-IC transition. 

\paragraph*{Conclusions}
In this Letter, we have studied two families of SU$(n)$ AKLT points.
In both cases, the AKLT points are characterized by kinks in the correlation lengths and square root singularities in the corresponding momenta.
The non-analytic behavior is explained by the coalescence of two eigenvalues of the transfer matrix, even though a single one is present precisely at the AKLT point. 
For the first family where the physical irrep is self-conjugate, we observe the conventional commensurate to incommensurate transition.
For the second family, the physical irrep is no longer self-conjugate, and the AKLT point is already incommensurate.
It remains singular, separating two incommensurate regimes within the same gapped phase.
The presence of a singularity within an incommensurate gapped phase represents a new phenomenon in quantum spin chains.

We conjecture that these conclusions apply to more complex translation-invariant AKLT states.
Preliminary results on more complex models seem to support the generality of these results, but the limitations of Clebsch-Gordan-based tensor networks prevent us from providing a definitive answer.
It would be interesting to generalize the approach of Ref.~\onlinecite{Nataf2018} to tensor algorithms, as they rely on the computation of the subduction coefficients only. 

More generally, these results show the strength of transfer matrix methods in infinite systems to characterize incommensurate phases. 
The transfer matrix generally remains an understudied property of quantum wavefunctions, in particular because of its pathological properties (non-Hermiticity, instabilities...), but the present results show that these difficulties can be overcome.

Finally, these results might have implications for classical systems, where disorder lines have been originally discussed. They suggest that singularities in the correlation length and in the wave vector could occur entirely inside an incommensurate phase. Which classical models could embody this possibility is left for future investigation.

\paragraph*{Acknowledgments.} We thank Lukas Devos, Olivier Gauthé, Keita Omiya, Pierre Nataf and Frank Schindler for insightful discussions. This work has been supported by the Swiss National Science Foundation Grants No. 212082 and 207202023001.

\bibliography{bibliography}

\iftrue
\appendix

\section{C-IC transition in the AKLT spin-1 chain.}\label{app:SU2}
The AKLT state\citep{AKLT, AKLT2} is a valence bond state where pairs of spin-$\frac{1}{2}$  on bonds form singlets.
Its parent Hamiltonian is
\begin{equation}
    H_{\mathrm{AKLT}} = \sum\limits \vec{T}_j \cdot \vec{T}_{j+1} + \beta \sum\limits (\vec{T}_j \cdot \vec{T}_{j+1})^2, \label{eq:parentSU2}
\end{equation}
with $\beta = \beta_\mathrm{AKLT} = \frac{1}{3}$ and $\vec{T}$ the generators of SU$(2)$.
For $ -1 < \beta < 1$\citep{Lauchli2006, DeChiara2011}, the system is in a topological phase with fractional spin-$\frac{1}{2}$ edge modes\citep{HaldaneConjecture1, HaldaneConjecture2, Kennedy1992, Lauchli2006, Pollmann2010}.
At the AKLT point, the exact AKLT state has bond dimension $2$, with correlation length $\frac{1}{\log 3}$.
As shown in Fig.~\ref{fig:SU2}, the spin-spin correlation are commensurate with momentum $\pi$ for $\beta < \beta_\mathrm{AKLT}$, and incommensurate for $\beta > \beta_{\mathrm{AKLT}}$.
\citep{Schollwock1996}.
Close to the AKLT point, the momentum scales as $q \propto \sqrt{\beta - \beta_\mathrm{AKLT}}$.
This is reminiscent of the scaling close to non-Hermitian exceptional points.\\

\begin{figure}
    \begin{center}
    \includegraphics[width=\linewidth]{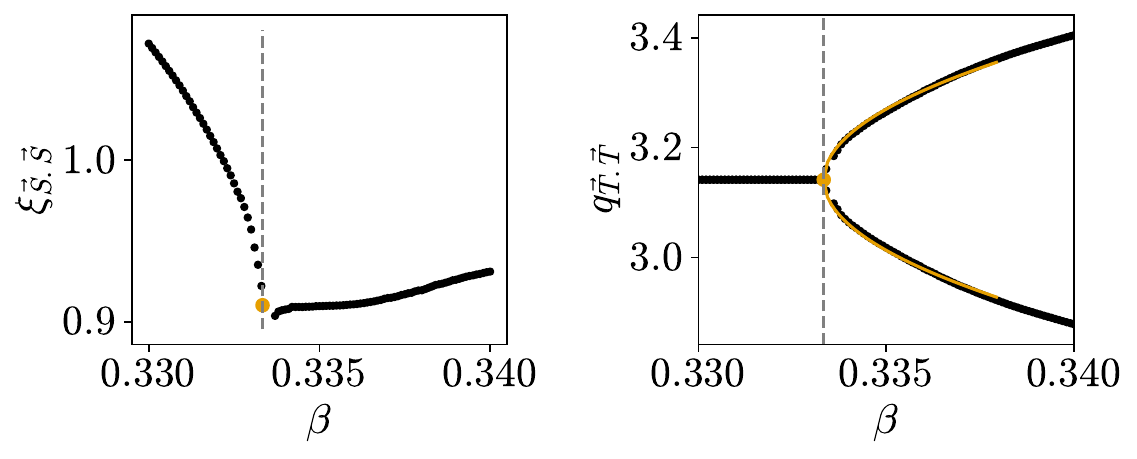}
    \end{center}
    \caption{Left: dominant correlation length associated to the spin-spin correlation. Right: the corresponding momentum. The orange line is a numerical fit of the form $\sqrt{\beta - \beta_\mathrm{AKLT}}$. The vertical dotted line marks the AKLT point. The C-IC transition is also characterized by  a cusp in the correlation length. The small oscillations of the correlation length close to the AKLT point illustrate the sensitivity of our computation to numerical noise due to the proximity of exceptional points.}
    \label{fig:SU2}
\end{figure}

\section{SU$(n)$ representation theory}\label{app:SU(n)}

In this section, we go over several elements of the representation theory of SU$(n)$.
Firstly, in Sec.~\ref{app:basics}, we introduce some notations to describe the irreps of SU$(n)$.
Then, in Sec.~\ref{app:parent}, we review the calculation of the parent Hamiltonians of several AKLT models considered in this paper.
Finally, in Sec.~\ref{app:TM}, we prove that the transfer matrix of a translation-invariant AKLT point is invariant if the physical irrep is self-conjugate.

\subsection{Basics of SU$(n)$ representations.}\label{app:basics}

The irreps of SU$(n)$ can be labelled by all the Young diagrams with strictly less than $n$ lines.
The fundamental representation is denoted by a simple box $\ydiagram{1}$.
The irreps $\ydiagram{2}$ (resp. $\ydiagram{3}$) is then the fully symmetric product of $2$ (resp. $3$) fundamental irreps, while $\ydiagram{1, 1}$ corresponds to an antisymmetric product of the two irreps.\\

The conjugate of an irrep corresponds to an irrep that transforms as if under the complex conjugate of the rotation.
Its Young diagram representation can be found as follows. 
Let $n_j$ be the number of columns with $j$ boxes.
$(n_1, ..., n_{N-1})$ fully characterize the irrep, and its conjugate is defined by $(n_{N-1}, ..., n_1)$.
Hence, for example, for SU$(3)$, the conjugate of the fundamental irrep is $\ydiagram{1, 1}$ and the conjugate of the three-box-symmetric irrep $\ydiagram{3}$ is $\ydiagram{3, 3}$.
If a Young diagram is left invariant under conjugation, the irrep is said to be self-conjugate.
The adjoint irrep with $n-1$ boxes in the first column and $1$ in the second is an example of such self-conjugate irreps.
Note that $SU(2)$ irreps are all self-conjugate.\\

Finally, to shorten notations, we also denote an irrep by its dimension: $\vb*{d}$.
For SU$(n \geq 3)$, pairs of conjugate irreps with the same dimension exist, and for SU$(n > 3)$, one even has irreps of the same dimension which are not conjugate to each other.
As such, the labelling is relatively arbitrary.
We will denote by $\vb*{\overline{d}}$ the conjugate of $\vb*{d}$.
In this paper, we use the following short-hand notations:
\begin{align}
    \vb*{1} &= \bullet \text{ the trivial irrep } &\text{ for SU$(n)$} \\
    \vb*{n} &= \ydiagram{1} &\text{ for SU$(n)$} \\
    \vb*{n^2 - 1} &= \text{ the adjoint } &\text{ for SU$(n)$} \\
    \vb*{8} &= \ydiagram{2, 1} \text{ (adjoint) } &\text{for SU$(3)$} \\
    \vb*{10} &= \ydiagram{3} &\text{ for SU$(3)$} \\
    \vb*{\overline{10}} &= \ydiagram{3, 3} &\text{ for SU$(3)$} \\
    \vb*{\overline{27}} &= \ydiagram{4, 2} &\text{ for SU$(3)$} \\
    \vb*{15} &= \ydiagram{2, 1, 1}  \text{ (adjoint) } &\text{for SU$(4)$}  \\
    \vb*{45} &= \ydiagram{3, 1} &\text{ for SU$(4)$} 
\end{align}

\subsection{Parent Hamiltonians for SU$(n)$ AKLT models.}\label{app:parent}
In this section, we summarize the derivation of the parent Hamiltonians of the studied AKLT chains. 
We start with the SU$(2)$ chain for reference.\\

\paragraph*{The SU$(2)$ spin-$1$ chain}
The local physical irrep is a spin 1, that is to say $\vb*{p} = \vb*{3}$, and the virtual legs are spin $\frac{1}{2}$.
From standard algebra, we know that $\vb*{3}\otimes \vb*{3} = \vb*{1} \oplus \vb*{3} \oplus\vb*{5}$, and $\vb*{2}\otimes \vb*{2} = \vb*{1} \oplus \vb*{3}$.
The Casimirs are given by 
\begin{align}
    C(\vb*{1}) = 0, \quad C(\vb*{3}) = 2, \quad C(\vb*{5}) = 6.
\end{align}
$h = \vec{T}^2(\vec{T}^2 - 2)$ is therefore an exact local Hamiltonian (it is already positive for $\vb*{5}$).
Developing $\vec{T} = \vec{T}_1 + \vec{T}_2$, we obtain that 
\begin{equation}
    \vec{T}^2 = 2 C(\vb*{3}) +  2\vec{T}_1\cdot\vec{T}_2,
\end{equation}
that is to say
\begin{equation}
    h =  (4 +  2\vec{T}_1\cdot\vec{T}_2) (2 +  2\vec{T}_1\cdot\vec{T}_2) = 8 + 12  \vec{T}_1\cdot\vec{T}_2 + 4(\vec{T}_1\cdot\vec{T}_2)^2.
\end{equation}
We recover, up to a global prefactor and a constant, the celebrated AKLT Hamiltonian.\\

\paragraph*{SU$(n)$ chains with self-conjugate representations}
We consider the states $\vert \vb*{n^2-1}; \vb*{\overline{n}}, \vb*{n} \rangle$.
For all values of $n$, we have 
\begin{equation}
    \vb*{n} \otimes \ \overline{\vb*{n}} = \bullet \oplus \vb*{n^2-1}.
\end{equation} 
Those two irreps also appear in the decomposition of $(\vb*{n^2-1})\otimes (\vb*{n^2-1})$.
The Casimir are $C(\bullet) = 0$ and $C(\vb*{n^2-1}) = n$, such that we can define
\begin{align}
    h &= \vec{T}^2 (\vec{T}^2 - n) \nonumber
    &= (2n + 2 \vec{T}_1\cdot\vec{T}_2) (n +  2\vec{T}_1\cdot\vec{T}_2).
\end{align}
Again, $h$ is already positive for all $n$.
We recover the Hamiltonian 
\begin{equation}
   H_{\mathbb{Z}_n} = \sum\limits \vec{T}_j.\vec{T}_{j+1} + \beta_\mathrm{AKLT} \sum\limits (\vec{T}_j.\vec{T}_{j+1})^2, \label{app:parentSUN}
\end{equation}
with $\beta_{\mathrm{AKLT}} = \frac{2}{3n}$.\\

\paragraph*{The 3-box symmetric AKLT chain}
The final model we consider is the SU$(3)$ AKLT state $\vert \vb*{10}; \vb*{8}, \vb*{8} \rangle$.
The only irreps in $\mathcal{I}$ are $\ydiagram{3, 3}$ and $\ydiagram{4, 2}$.
Their Casimir are $6$ and $8$, such that we define 
\begin{align}
    h &= (\vec{T}^2 - 6) (\vec{T}^2 - 8) \\
    &= (6 + 2 \vec{T}_1.\vec{T}_2) (4 + 2\vec{T}_1.\vec{T}_2).
\end{align}
It is already positive definite (all other relevant irreps have a Casimir smaller than $6$), and we recover the parent Hamiltonian.

\subsection{Transfer matrix at the AKLT point}\label{app:TM}
We consider an AKLT point defined as in the main text by:
\begin{equation}
    \localtensor
\end{equation}
chosen such that $\ket{\vb*{p} ; \vb*{v_L}, \vb*{v_R}}$ is normalized.
In these notations, one can consider $\vb*{v_L}$ and $ \vb*{v_R}$ as input indices of the tensor and $\vb*{p}$ as its output (or the other way around).
To analyse the transfer matrix, it is more convenient to work with a marginally different convention, where the leg $\vb*{v_R}$ is taken as an output leg (and therefore transforms as $\vb*{\overline{v}_R}$)  and the other two as inputs.
The tensor $M$ can then be written as
\begin{equation}
    M = \sum\limits_{a, \alpha, b} M_{a, \alpha}^{b} \vert \vb*{v_L}, a \rangle \vert \vb*{p}, \alpha \rangle \langle \vb*{\overline{v}_R}, b \vert,
\end{equation}
where $(a, \alpha, b)$ label the different bases.
The Wickner-Eckart theorem enforces that 
\begin{equation}
    M = \mathcal{M} \sum\limits_{a, \alpha, b} C_{\vb*{v_L}, a; \vb*{p}, \alpha}^{\vb*{\overline{v}_R}, b} \vert \vb*{v_L}, a \rangle \vert \vb*{p}, \alpha \rangle \langle \vb*{\overline{v}_R}, b \vert,
\end{equation}
where $C_{\vb*{v_L}, a; \vb*{p}, \alpha}^{\vb*{\overline{v}_R}, b} $ is the real Clebsch-Gordan coefficient defined by
\begin{equation}
    C_{\vb*{v_L}, a; \vb*{p}, \alpha}^{\vb*{\overline{v}_R}, b} =  \langle \vb*{v_L}, a ;  \vb*{p}, \alpha \vert  \vb*{\overline{v}_R}, b \rangle
\end{equation}
and $\mathcal{M}$ is a complex constant.
We take $\mathcal{M}$ of modulus $1$ such that $M$ is left- (or right-) unitary.\\

For the AKLT states invariant under translation by one site that we consider in this paper, $\vb*{\overline{v}_R} = \vb*{v_L} \equiv \vb*{v}$, such that
\begin{equation}
    M = \mathcal{M} \sum\limits_{a, \alpha, b} C_{\vb*{v}, a; \vb*{p}, \alpha}^{\vb*v, b} \vert \vb*v, a \rangle \vert \vb*{p}, \alpha \rangle \langle \vb*v, b \vert.
\end{equation}
The corresponding transfer matrix is straightforwardly derived to be:
\begin{multline}
    T_M = \sum\limits_{a_T, a_B; b_T b_B } \sum\limits_{\alpha} C_{\vb*{v}, a_T; \vb*{p}, \alpha}^{\vb*v, b_T} C_{\vb*{v}, a_B; \vb*{p}, \alpha}^{\vb*v, b_B}\\
    \vert \vb*v, a_T ;  \vb*{\overline{v}}, \overline{a}_B \rangle \langle \vb*v, b_T ; \vb*{\overline{v}}, \overline{b}_B \vert.
\end{multline}
The irreps $\vb*{\overline{v}}$ correspond to the bottom legs of the transfer matrices.
We denote $\overline{a}$ in the irrep $\vb*{\overline{v}}$ the conjugate of the state $a$ in the irrep $\vb*v$.\\

At the AKLT point, we can guarantee that the system is commensurate if $T_M$ is Hermitian, that is to say if
\begin{equation}
    \sum\limits_{\alpha} C_{\vb*{v}, a_T; \vb*{p}, \alpha}^{\vb*v, b_T} C_{\vb*{v}, a_B; \vb*{p}, \alpha}^{\vb*v, b_B} = \sum\limits_{\alpha} C_{\vb*{v}, b_T; \vb*{p}, \alpha}^{\vb*v, a_T} C_{\vb*{v}, b_B; \vb*{p}, \alpha}^{\vb*v, a_B}.
\end{equation}
This equality is not trivial, as the Clebsch-Gordan coefficients only verify:
\begin{equation}
     C_{\vb*{v}, a; \vb*{p}, \alpha}^{\vb*v, b} = \xi(\vb*{p}, \alpha)  C_{\vb*{v}, b; \vb*{\overline{p}}, \overline{\alpha}}^{\vb*{v}, a}, \label{eq:Lemma}
\end{equation}
where $\xi = \pm 1$ depends only on $\vb*{p}$ and $\alpha$ (proof below).
Nonetheless, if the irrep is self-conjugate, we can check that 
\begin{align}
    \sum\limits_{\alpha} C_{\vb*{v}, a_T; \vb*{p}, \alpha}^{\vb*v, b_T} C_{\vb*{v}, a_B; \vb*{p}, \alpha}^{\vb*v, b_B} &= \sum\limits_{\alpha} C_{\vb*{v}, b_T; \vb*{p}, \overline{\alpha}}^{\vb*v, a_T} C_{\vb*{v}, b_B; \vb*{p}, \overline{\alpha}}^{\vb*v, a_B} \xi(\vb*{p}, \alpha)^2 \\
    &= \sum\limits_{\overline{\alpha}} C_{\vb*{v}, b_T; \vb*{p}, \overline{\alpha}}^{\vb*v, a_T} C_{\vb*{v}, b_B; \vb*{p}, \overline{\alpha}}^{\vb*v, a_B},
\end{align}
and $T_M$ is Hermitian.\\

\paragraph*{Proof of Eq.~\eqref{eq:Lemma}:}
We define the SU$(n)$ invariant tensor $P$ by 
\begin{align}
    P &= \frac{1}{C_{\vb*{p}, 1 ; \vb*{\overline{p}} , \overline{1}}^{\vb*1, 1}} \sum\limits_{\alpha, \beta} C_{\vb*{p}, \alpha ; \vb*{\overline{p}} , \beta}^{\vb*1, 1} \vert \vb*1, 1 \rangle \langle \vb*p , \alpha ; \vb*{\overline{p}} , \beta \vert \\
    &= \sum\limits_{\alpha} \xi(\vb*p, \alpha) \vert \vb*1, 1 \rangle \langle \vb*p , \alpha ; \vb*{\overline{p}} , \overline{\alpha} \vert.
\end{align}
The trivial representation $\vb*1$ is of dimension $1$ and can be ignored in tensor calculations.
Contracting $M$ and $P$ leaves us with the tensor
\begin{equation}
    MP = \mathcal{M} \sum\limits_{a, \alpha, b} C_{\vb*{v}, a; \vb*{p}, \alpha}^{\vb*v, b} \xi(\vb*p, \alpha) \vert \vb*v, a\rangle  \langle \vb*v, b ; \vb*{\overline{p}} , \overline{\alpha} \vert,
\end{equation}
which should be equal, according to the Wigner-Eckart theorem, to
\begin{equation}
    MP = \tilde{\mathcal{M}} \sum\limits_{a, \alpha, b} C_{\vb*{v}, b; \vb*{\overline{p}} , \overline{\alpha}}^{\vb*v, a}  \vert \vb*v, a \rangle   \langle \vb*v, b ; \vb*{\overline{p}} , \overline{\alpha} \vert.
\end{equation}
Given that all states and Clebsch-Gordan coefficients are real, up to a redefinition of $\xi$, we obtain the equality Eq.~\eqref{eq:Lemma} by basis identification.

\section{Additional numerical results}\label{app:AddData}
In this section, we provide some additional numerical results to complement the main text.

%
%


Firtly, we represent in Fig.~\ref{fig:SU5} the correlation length and the momentum of the spin-spin correlation for the SU$(5)$ AKLT chain in the adjoint representation.
Though our data is limited by the computation of the Clebsch-Gordan coefficients of the large SU$(5)$ irreps, the AKLT point clearly remains a disorder point.

\begin{figure}
    \centering
    \includegraphics[width=0.49\linewidth]{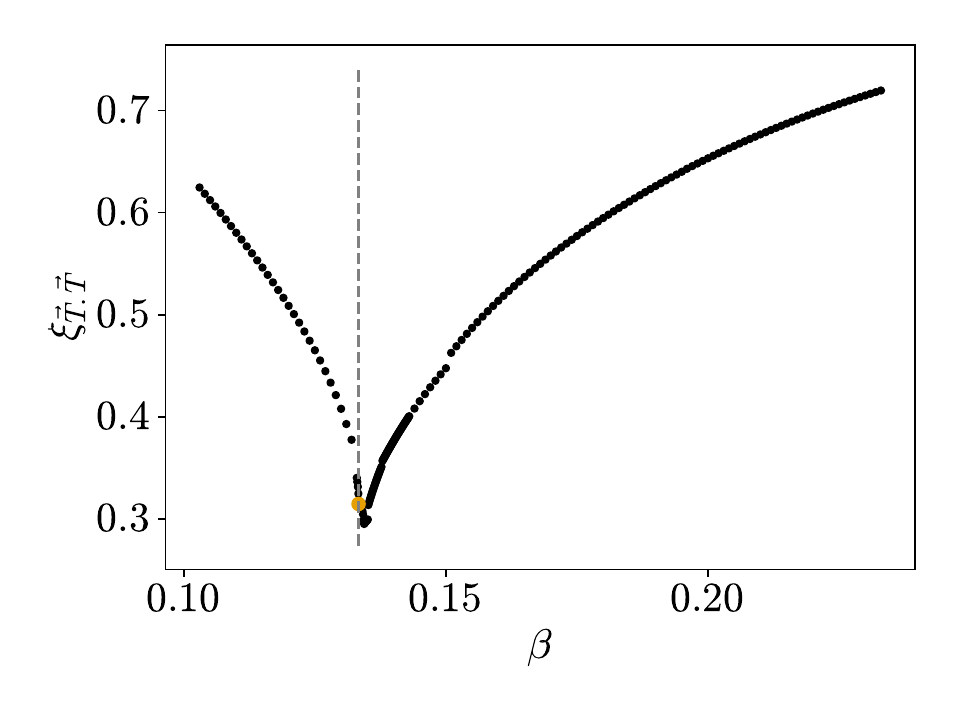}
    \includegraphics[width=0.49\linewidth]{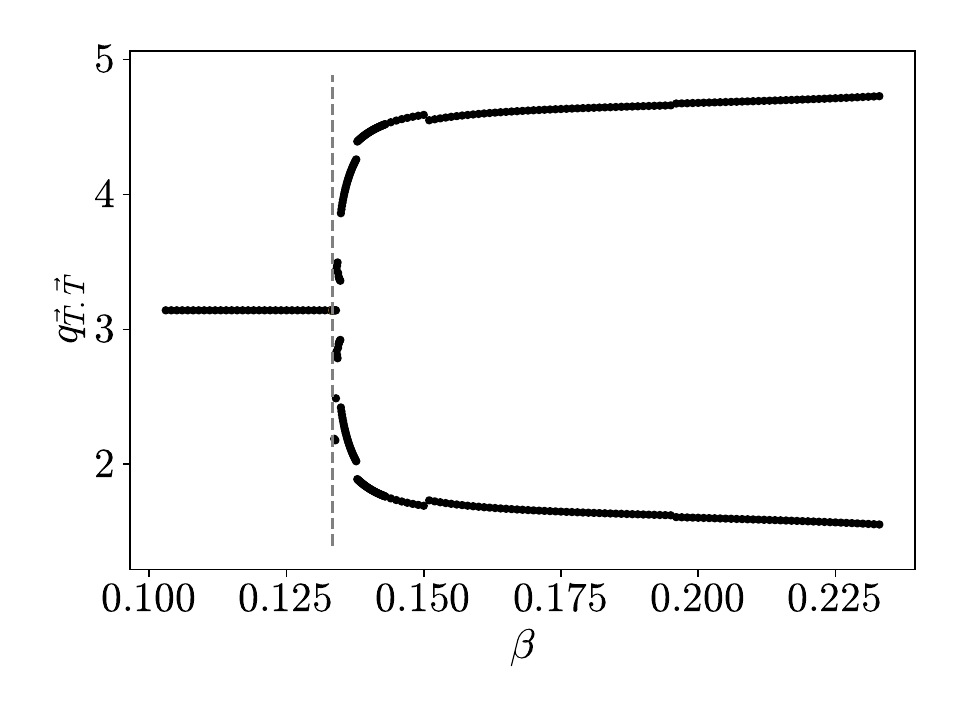}
    \caption{Left: the largest correlation length associated to the spin-spin correlations following Eq.~\eqref{eq:recipe} for SU$(5)$. Bottom: momentum associated to the leading eigenvalues. The dashed line marks the AKLT point. }
    \label{fig:SU5}
\end{figure}

Furthermore, we show in Fig.~\ref{fig:SU3-all} the correlation lengths in the adjoint irrep of the SU$(3)$ self-conjugate model.
As discussed in the main text, a large number of unphysical eigenvalues appears.
While there is significant variation in the spectrum of the transfer matrix, the iMPS evolve smoothly when varying $\beta$.
The small components of the wavefunctions contribute significantly to the spectrum of the transfer matrix as the weights (the entanglement spectrum) do not explicitly intervene in the left- (or right-) transfer matrix.
To extract the physical information, we have to select the correct contributions by taking into account the overlaps with physically relevant environments.
\begin{figure}
    \centering
    \includegraphics[width=0.89\linewidth]{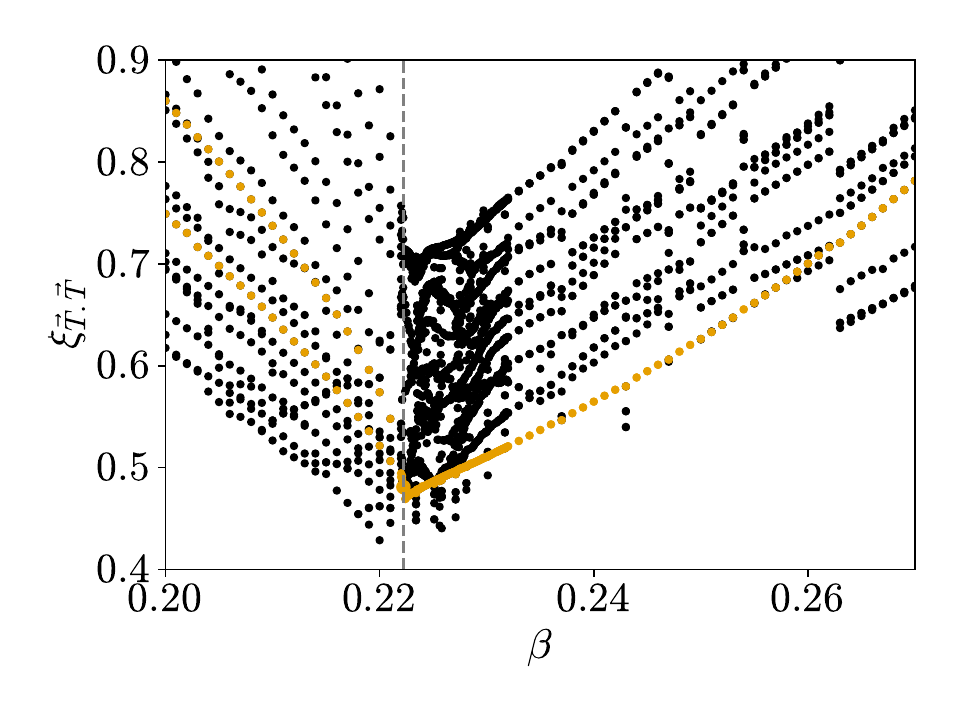}
    \caption{Largest correlation lengths  in the adjoint sector extracted from the transfer matrix of the SU$(3)$ fundamental model. In orange, we underline the physically relevant ones. The large number of unphysical eigenvalues arise from the very short correlation length and numerical instabilities.}
    \label{fig:SU3-all}
\end{figure}

Finally, in Fig~\ref{fig:qlowbeta}, we present the momentum of the spin-spin correlation for $\beta < \beta_\mathrm{AKLT}$.
We observe a slow, approximately gaussian decay of the momentum towards the commensurate value $\frac{2\pi}{3}$.
This confirms the picture of an incommensurate phase (out of $\beta = 0$).
The incertitude on the exact value of the momentum for $\beta \leq 0.05$ is not negligible, despite a large bond dimension, but the correlations remain convincingly incommensurate well below $\beta_\mathrm{class}$.

\begin{figure}
    \centering
    \includegraphics[width=0.9\linewidth]{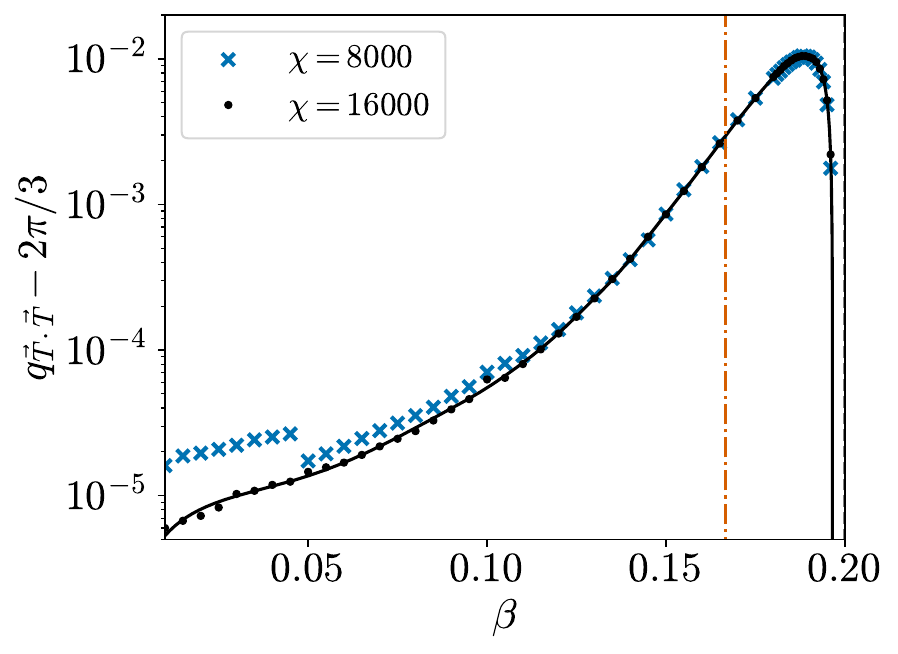}
    \caption{Behavior of the momentum $q_{\vec{T}\cdot \vec{T}}$ in the three box-symmetric model for $\beta < \beta_{AKLT}$. While the precision on the momentum is limited for $\beta < 0.05$, it remains convincingly incommensurate well below $\beta_\mathrm{class}$ (the orange dashed line). }
    \label{fig:qlowbeta}
\end{figure}

\section{Semiclassical approach}\label{app:classical}
We follow the notations and approach of Refs.~\onlinecite{Lajko2017, Wamer2020} to describe the SU($3$) 3-box symmetric model.
We introduce the operators $\tilde{S}_b^a$, which are matrices of dimension $(n \times n)$ and of trace $p = 3$.
The relations to standard spin operators is therefore $S_b^a = \tilde{S}_b^a - \frac{p}{n} \delta_b^a$.
We introduce the fields
\begin{equation}
    \tilde{S}_b^a = p \overline{\phi}^a \phi_b.
\end{equation}
For classical fields, we take $\vert \phi \vert = 1$.
It is straightforward to compute:
\begin{eqnarray}
    \vec{S}(m).\vec{S}(n) &=  S_b^a(m) S_a^b(n) =  \tilde{S}_b^a(m) \tilde{S}_a^b(n) - \frac{p}{n} \\
    &=  p^2 \overline{\phi}^a(m) \phi_b(m) \overline{\phi}^b(n) \phi_a(n) - \frac{p^2}{n} \\
     &=  p^2 \vert \overline{\phi}(m). \phi(n)\vert^2 - \frac{p^2}{n} 
\end{eqnarray}
This computation is the starting point for the flavourwave expansion, with many degenerate configurations.
The typical way to cure the degeneracy to recover the classical limit of the quantum GS is to work with an effective model
\begin{multline}
    H = \sum\limits_j J_1  \vec{S}(j).\vec{S}(j+1) + J_2  \vec{S}(j).\vec{S}(j+2) \\
    -J_3  \vec{S}(j).\vec{S}(j+3),
\end{multline}
with all $J$s positive.\\

Let us forgo for now the additional couplings. 
The biquadratic term is simply
\begin{multline}
    \left[ \vec{S}(m).\vec{S}(n) \right]^2 = p^4 \vert \overline{\phi}(m). \phi(n)\vert^4  
    \\ -   \frac{2p^4}{n} \vert \overline{\phi}(m). \phi(n)\vert^2 +   \frac{p^4}{n^2} 
\end{multline}

Then, our Hamiltonian can be written, up to an unimportant constant, as
\begin{equation}
    \frac{H}{p^2} = \sum\limits_j \beta p^2 \vert \overline{\phi}(j). \phi(j+1)\vert^4 + (1 -  \frac{2 \beta p^2}{n}) \vert \overline{\phi}(j). \phi(j+1)\vert^2
\end{equation}
This is minimal for 
\begin{equation}
    \vert \overline{\phi}(j). \phi(j+1)\vert^2 = \frac{2 \beta p^2 - n}{2 \beta p^2}.
\end{equation}
As long as $\beta < \frac{n}{2 p^2}$, the classical groundstate is therefore unchanged.
For  $\beta > \frac{n}{2 p^2}$, the classical groundstate is a plane wave with incommensurate momentum.
The momentum scales as $\sqrt{\beta - \beta_c}$ as expected. \\ 

Introducing $J_2$ and $J_3$ does not change this picture: the classical groundstate is commensurate up to a critical value of $\beta$.

\fi
\end{document}

%% file: tikzfig.tex
\newcommand{\tensorwidth}{0.6cm} 
\newcommand{\tensorheight}{0.6cm} 

\def\transfermatrix{\tikz[baseline=.1ex]{
\% Nodes with controllable width and height for upper row
        \node[draw, minimum width=\tensorwidth, minimum height=\tensorheight] (A11) at (0, 0.5) {$L_{1}$};
        \node[draw, minimum width=\tensorwidth, minimum height=\tensorheight] (A12) at (0.9, 0.5) {$L_{2}$};

        \node[draw, minimum width=\tensorwidth, minimum height=\tensorheight] (B11) at (0, -0.5) {$L_{1}^\dagger$};
        \node[draw, minimum width=\tensorwidth, minimum height=\tensorheight] (B12) at (0.9, -0.5) {$L_{2}^\dagger$};

        \draw (A11.east) -- (A12.west);
        \draw (A11.west) -- ++(-0.25,0);
        \draw (A12.east) -- ++(0.25,0);

        \draw (B11.east) -- (B12.west);
        \draw (B11.west) -- ++(-0.25,0);
        \draw (B12.east) -- ++(0.25,0);

        \draw (A11.south) -- (B11.north);
        \draw (A12.south) -- (B12.north);
       
}
}

\def\bigtransfermatrix{\tikz[baseline=.1ex]{
\% Nodes with controllable width and height for upper row
        \node[draw, minimum width=\tensorwidth, minimum height=\tensorheight] (A11) at (0, 1) {$L_{1}$};
        \node[draw, minimum width=\tensorwidth, minimum height=\tensorheight] (A12) at (0.9, 1) {$L_{2}$};

        \node[draw, minimum width=\tensorwidth, minimum height=\tensorheight] (B11) at (0, -1) {$L_{1}^\dagger$};
        \node[draw, minimum width=\tensorwidth, minimum height=\tensorheight] (B12) at (0.9, -1) {$L_{2}^\dagger$};

        \draw (A11.east) -- (A12.west);
        \draw (A11.west) -- ++(-0.25,0);
        \draw (A12.east) -- ++(0.25,0);

        \draw (B11.east) -- (B12.west);
        \draw (B11.west) -- ++(-0.25,0);
        \draw (B12.east) -- ++(0.25,0);

        \draw (A11.south) -- (B11.north);
        \draw (A12.south) -- (B12.north);
}
}

\def\leftvector{\tikz[baseline=.1ex]{
\% Nodes with controllable width and height for upper row
        \node[draw, minimum width=\tensorwidth, minimum height=\tensorheight] (A1) at (0, 1) {$L_{2}$};

        \node[draw, minimum width=\tensorwidth, minimum height=\tensorheight] (O) at (0, 0) {$\vec{O}_x$};
        
        \node[draw, minimum width=\tensorwidth, minimum height=\tensorheight] (B1) at (0, -1) {$L_{2}^\dagger$};

        \draw (A1.west) -- ++(-0.25,0) -- ++(0, -2);
        \draw (A1.east) -- ++(0.25,0);
        \draw (B1.west) -- ++(-0.25,0);
        \draw (B1.east) -- ++(0.25,0);
        \draw (O.east) -- ++(0.25,0);

        \draw (A1.south) -- (O.north);
        \draw (B1.north) -- (O.south);

}
}

\def\rightvector{\tikz[baseline=.1ex]{
\% Nodes with controllable width and height for upper row
        \node[draw, minimum width=\tensorwidth, minimum height=\tensorheight] (A1) at (0, 1) {$L_{1}$};
        \node[draw, minimum width=\tensorwidth, minimum height=\tensorheight] (A2) at (0.9, 1) {$L_{2}$};
        \node[draw, minimum width=\tensorwidth, minimum height=\tensorheight] (AC) at (1.8, 1) {$C$};

        \node[draw, minimum width=\tensorwidth, minimum height=\tensorheight] (O) at (0, 0) {$\vec{O}_y$};
        
        \node[draw, minimum width=\tensorwidth, minimum height=\tensorheight] (B1) at (0, -1) {$L_{1}^\dagger$};
        \node[draw, minimum width=\tensorwidth, minimum height=\tensorheight] (B2) at (0.9, -1) {$L_{2}^\dagger$};
        \node[draw, minimum width=\tensorwidth, minimum height=\tensorheight] (BC) at (1.8, -1) {$C^\dagger$};
        
        \draw (A1.east) -- (A2.west);
        \draw (A2.east) -- (AC.west);
        \draw (A1.west) -- ++(-0.25,0);
        \draw (AC.east) -- ++(0.25,0) -- ++(0, -2);
        \draw (B1.east) -- (B2.west);
        \draw (B2.east) -- (BC.west);
        \draw (B1.west) -- ++(-0.25,0);
        \draw (BC.east) -- ++(0.25,0);
        \draw (O.west) -- ++(-0.25,0);

        \draw (A1.south) -- (O.north);
        \draw (A2.south) -- (B2.north);
        \draw (B1.north) -- (O.south);
       
}
}

\def\localtensor{\tikz{
\% Nodes with controllable width and height for upper row
        \node[draw, minimum width=\tensorwidth, minimum height=\tensorheight] (M) at (0, 1) {$M$};

        \draw (M.west) -- ++(-0.25,0) node [label=left:{$\vb*{v_L}$}] {};
        \draw (M.east) -- ++(0.25,0) node [label=right:{$\vb*{v_R}$}] {};
        \draw (M.south) -- ++(0,-0.25) node [label=south:{$\vb*{p}$}] {};

}
}

%% file: main.bbl
\begin{thebibliography}{40}%
\makeatletter
\providecommand \@ifxundefined [1]{%
 \@ifx{#1\undefined}
}%
\providecommand \@ifnum [1]{%
 \ifnum #1\expandafter \@firstoftwo
 \else \expandafter \@secondoftwo
 \fi
}%
\providecommand \@ifx [1]{%
 \ifx #1\expandafter \@firstoftwo
 \else \expandafter \@secondoftwo
 \fi
}%
\providecommand \natexlab [1]{#1}%
\providecommand \enquote  [1]{``#1''}%
\providecommand \bibnamefont  [1]{#1}%
\providecommand \bibfnamefont [1]{#1}%
\providecommand \citenamefont [1]{#1}%
\providecommand \href@noop [0]{\@secondoftwo}%
\providecommand \href [0]{\begingroup \@sanitize@url \@href}%
\providecommand \@href[1]{\@@startlink{#1}\@@href}%
\providecommand \@@href[1]{\endgroup#1\@@endlink}%
\providecommand \@sanitize@url [0]{\catcode `\\12\catcode `\$12\catcode `\&12\catcode `\#12\catcode `\^12\catcode `\_12\catcode `\%12\relax}%
\providecommand \@@startlink[1]{}%
\providecommand \@@endlink[0]{}%
\providecommand \url  [0]{\begingroup\@sanitize@url \@url }%
\providecommand \@url [1]{\endgroup\@href {#1}{\urlprefix }}%
\providecommand \urlprefix  [0]{URL }%
\providecommand \Eprint [0]{\href }%
\providecommand \doibase [0]{http://dx.doi.org/}%
\providecommand \selectlanguage [0]{\@gobble}%
\providecommand \bibinfo  [0]{\@secondoftwo}%
\providecommand \bibfield  [0]{\@secondoftwo}%
\providecommand \translation [1]{[#1]}%
\providecommand \BibitemOpen [0]{}%
\providecommand \bibitemStop [0]{}%
\providecommand \bibitemNoStop [0]{.\EOS\space}%
\providecommand \EOS [0]{\spacefactor3000\relax}%
\providecommand \BibitemShut  [1]{\csname bibitem#1\endcsname}%
\let\auto@bib@innerbib\@empty
\bibitem [{\citenamefont {Affleck}\ \emph {et~al.}(1987)\citenamefont {Affleck}, \citenamefont {Kennedy}, \citenamefont {Lieb},\ and\ \citenamefont {Tasaki}}]{AKLT}%
  \BibitemOpen
  \bibfield  {author} {\bibinfo {author} {\bibfnamefont {Ian}\ \bibnamefont {Affleck}}, \bibinfo {author} {\bibfnamefont {Tom}\ \bibnamefont {Kennedy}}, \bibinfo {author} {\bibfnamefont {Elliott~H.}\ \bibnamefont {Lieb}}, \ and\ \bibinfo {author} {\bibfnamefont {Hal}\ \bibnamefont {Tasaki}},\ }\bibfield  {title} {\enquote {\bibinfo {title} {Rigorous results on valence-bond ground states in antiferromagnets},}\ }\href {\doibase 10.1103/PhysRevLett.59.799} {\bibfield  {journal} {\bibinfo  {journal} {Phys. Rev. Lett.}\ }\textbf {\bibinfo {volume} {59}},\ \bibinfo {pages} {799--802} (\bibinfo {year} {1987})}\BibitemShut {NoStop}%
\bibitem [{\citenamefont {Haldane}(1983{\natexlab{a}})}]{HaldaneConjecture1}%
  \BibitemOpen
  \bibfield  {author} {\bibinfo {author} {\bibfnamefont {F.~D.~M.}\ \bibnamefont {Haldane}},\ }\bibfield  {title} {\enquote {\bibinfo {title} {Nonlinear field theory of large-spin heisenberg antiferromagnets: Semiclassically quantized solitons of the one-dimensional easy-axis n\'eel state},}\ }\href {\doibase 10.1103/PhysRevLett.50.1153} {\bibfield  {journal} {\bibinfo  {journal} {Phys. Rev. Lett.}\ }\textbf {\bibinfo {volume} {50}},\ \bibinfo {pages} {1153--1156} (\bibinfo {year} {1983}{\natexlab{a}})}\BibitemShut {NoStop}%
\bibitem [{\citenamefont {Haldane}(1983{\natexlab{b}})}]{HaldaneConjecture2}%
  \BibitemOpen
  \bibfield  {author} {\bibinfo {author} {\bibfnamefont {F.D.M.}\ \bibnamefont {Haldane}},\ }\bibfield  {title} {\enquote {\bibinfo {title} {Continuum dynamics of the 1-d heisenberg antiferromagnet: Identification with the o(3) nonlinear sigma model},}\ }\href {\doibase https://doi.org/10.1016/0375-9601(83)90631-X} {\bibfield  {journal} {\bibinfo  {journal} {Physics Letters A}\ }\textbf {\bibinfo {volume} {93}},\ \bibinfo {pages} {464--468} (\bibinfo {year} {1983}{\natexlab{b}})}\BibitemShut {NoStop}%
\bibitem [{\citenamefont {Affleck}\ \emph {et~al.}(1988)\citenamefont {Affleck}, \citenamefont {Kennedy}, \citenamefont {Lieb},\ and\ \citenamefont {Tasaki}}]{AKLT2}%
  \BibitemOpen
  \bibfield  {author} {\bibinfo {author} {\bibfnamefont {Ian}\ \bibnamefont {Affleck}}, \bibinfo {author} {\bibfnamefont {Tom}\ \bibnamefont {Kennedy}}, \bibinfo {author} {\bibfnamefont {Elliott~H.}\ \bibnamefont {Lieb}}, \ and\ \bibinfo {author} {\bibfnamefont {Hal}\ \bibnamefont {Tasaki}},\ }\bibfield  {title} {\enquote {\bibinfo {title} {Valence bond ground states in isotropic quantum antiferromagnets},}\ }\href {\doibase 10.1007/BF01218021} {\bibfield  {journal} {\bibinfo  {journal} {Communications in Mathematical Physics}\ }\textbf {\bibinfo {volume} {115}},\ \bibinfo {pages} {477--528} (\bibinfo {year} {1988})}\BibitemShut {NoStop}%
\bibitem [{\citenamefont {Schollwöck}(2011)}]{SchollwockDMRG}%
  \BibitemOpen
  \bibfield  {author} {\bibinfo {author} {\bibfnamefont {Ulrich}\ \bibnamefont {Schollwöck}},\ }\bibfield  {title} {\enquote {\bibinfo {title} {The density-matrix renormalization group in the age of matrix product states},}\ }\href {\doibase https://doi.org/10.1016/j.aop.2010.09.012} {\bibfield  {journal} {\bibinfo  {journal} {Annals of Physics}\ }\textbf {\bibinfo {volume} {326}},\ \bibinfo {pages} {96--192} (\bibinfo {year} {2011})},\ \bibinfo {note} {january 2011 Special Issue}\BibitemShut {NoStop}%
\bibitem [{\citenamefont {L\"auchli}\ \emph {et~al.}(2006)\citenamefont {L\"auchli}, \citenamefont {Schmid},\ and\ \citenamefont {Trebst}}]{Lauchli2006}%
  \BibitemOpen
  \bibfield  {author} {\bibinfo {author} {\bibfnamefont {Andreas}\ \bibnamefont {L\"auchli}}, \bibinfo {author} {\bibfnamefont {Guido}\ \bibnamefont {Schmid}}, \ and\ \bibinfo {author} {\bibfnamefont {Simon}\ \bibnamefont {Trebst}},\ }\bibfield  {title} {\enquote {\bibinfo {title} {Spin nematics correlations in bilinear-biquadratic $s=1$ spin chains},}\ }\href {\doibase 10.1103/PhysRevB.74.144426} {\bibfield  {journal} {\bibinfo  {journal} {Phys. Rev. B}\ }\textbf {\bibinfo {volume} {74}},\ \bibinfo {pages} {144426} (\bibinfo {year} {2006})}\BibitemShut {NoStop}%
\bibitem [{\citenamefont {De~Chiara}\ \emph {et~al.}(2011)\citenamefont {De~Chiara}, \citenamefont {Lewenstein},\ and\ \citenamefont {Sanpera}}]{DeChiara2011}%
  \BibitemOpen
  \bibfield  {author} {\bibinfo {author} {\bibfnamefont {G.}~\bibnamefont {De~Chiara}}, \bibinfo {author} {\bibfnamefont {M.}~\bibnamefont {Lewenstein}}, \ and\ \bibinfo {author} {\bibfnamefont {A.}~\bibnamefont {Sanpera}},\ }\bibfield  {title} {\enquote {\bibinfo {title} {Bilinear-biquadratic spin-1 chain undergoing quadratic zeeman effect},}\ }\href {\doibase 10.1103/PhysRevB.84.054451} {\bibfield  {journal} {\bibinfo  {journal} {Phys. Rev. B}\ }\textbf {\bibinfo {volume} {84}},\ \bibinfo {pages} {054451} (\bibinfo {year} {2011})}\BibitemShut {NoStop}%
\bibitem [{\citenamefont {Kennedy}\ and\ \citenamefont {Tasaki}(1992)}]{Kennedy1992}%
  \BibitemOpen
  \bibfield  {author} {\bibinfo {author} {\bibfnamefont {Tom}\ \bibnamefont {Kennedy}}\ and\ \bibinfo {author} {\bibfnamefont {Hal}\ \bibnamefont {Tasaki}},\ }\bibfield  {title} {\enquote {\bibinfo {title} {Hidden ${\mathrm{z}}_{2}$\ifmmode\times\else\texttimes\fi{}${\mathrm{z}}_{2}$ symmetry breaking in haldane-gap antiferromagnets},}\ }\href {\doibase 10.1103/PhysRevB.45.304} {\bibfield  {journal} {\bibinfo  {journal} {Phys. Rev. B}\ }\textbf {\bibinfo {volume} {45}},\ \bibinfo {pages} {304--307} (\bibinfo {year} {1992})}\BibitemShut {NoStop}%
\bibitem [{\citenamefont {Pollmann}\ \emph {et~al.}(2010)\citenamefont {Pollmann}, \citenamefont {Turner}, \citenamefont {Berg},\ and\ \citenamefont {Oshikawa}}]{Pollmann2010}%
  \BibitemOpen
  \bibfield  {author} {\bibinfo {author} {\bibfnamefont {Frank}\ \bibnamefont {Pollmann}}, \bibinfo {author} {\bibfnamefont {Ari~M.}\ \bibnamefont {Turner}}, \bibinfo {author} {\bibfnamefont {Erez}\ \bibnamefont {Berg}}, \ and\ \bibinfo {author} {\bibfnamefont {Masaki}\ \bibnamefont {Oshikawa}},\ }\bibfield  {title} {\enquote {\bibinfo {title} {Entanglement spectrum of a topological phase in one dimension},}\ }\href {\doibase 10.1103/PhysRevB.81.064439} {\bibfield  {journal} {\bibinfo  {journal} {Phys. Rev. B}\ }\textbf {\bibinfo {volume} {81}},\ \bibinfo {pages} {064439} (\bibinfo {year} {2010})}\BibitemShut {NoStop}%
\bibitem [{\citenamefont {Auerbach}(1994)}]{book_auerbach}%
  \BibitemOpen
  \bibfield  {author} {\bibinfo {author} {\bibfnamefont {Assa}\ \bibnamefont {Auerbach}},\ }\href@noop {} {\emph {\bibinfo {title} {Interacting Electrons and Quantum Magnetism}}}\ (\bibinfo  {publisher} {Springer-Verlag},\ \bibinfo {year} {1994})\BibitemShut {NoStop}%
\bibitem [{\citenamefont {Verstraete}\ and\ \citenamefont {Cirac}(2004{\natexlab{a}})}]{verstraete1}%
  \BibitemOpen
  \bibfield  {author} {\bibinfo {author} {\bibfnamefont {F.}~\bibnamefont {Verstraete}}\ and\ \bibinfo {author} {\bibfnamefont {J.~I.}\ \bibnamefont {Cirac}},\ }\href {https://arxiv.org/abs/cond-mat/0407066} {\enquote {\bibinfo {title} {Renormalization algorithms for quantum-many body systems in two and higher dimensions},}\ } (\bibinfo {year} {2004}{\natexlab{a}}),\ \Eprint {http://arxiv.org/abs/cond-mat/0407066} {arXiv:cond-mat/0407066 [cond-mat.str-el]} \BibitemShut {NoStop}%
\bibitem [{\citenamefont {Verstraete}\ and\ \citenamefont {Cirac}(2004{\natexlab{b}})}]{verstraete2}%
  \BibitemOpen
  \bibfield  {author} {\bibinfo {author} {\bibfnamefont {F.}~\bibnamefont {Verstraete}}\ and\ \bibinfo {author} {\bibfnamefont {J.~I.}\ \bibnamefont {Cirac}},\ }\bibfield  {title} {\enquote {\bibinfo {title} {Valence-bond states for quantum computation},}\ }\href {\doibase 10.1103/PhysRevA.70.060302} {\bibfield  {journal} {\bibinfo  {journal} {Phys. Rev. A}\ }\textbf {\bibinfo {volume} {70}},\ \bibinfo {pages} {060302} (\bibinfo {year} {2004}{\natexlab{b}})}\BibitemShut {NoStop}%
\bibitem [{\citenamefont {Schollw\"ock}\ \emph {et~al.}(1996)\citenamefont {Schollw\"ock}, \citenamefont {Jolic\oe{}ur},\ and\ \citenamefont {Garel}}]{Schollwock1996}%
  \BibitemOpen
  \bibfield  {author} {\bibinfo {author} {\bibfnamefont {U.}~\bibnamefont {Schollw\"ock}}, \bibinfo {author} {\bibfnamefont {Th.}\ \bibnamefont {Jolic\oe{}ur}}, \ and\ \bibinfo {author} {\bibfnamefont {T.}~\bibnamefont {Garel}},\ }\bibfield  {title} {\enquote {\bibinfo {title} {Onset of incommensurability at the valence-bond-solid point in the s=1 quantum spin chain},}\ }\href {\doibase 10.1103/PhysRevB.53.3304} {\bibfield  {journal} {\bibinfo  {journal} {Phys. Rev. B}\ }\textbf {\bibinfo {volume} {53}},\ \bibinfo {pages} {3304--3311} (\bibinfo {year} {1996})}\BibitemShut {NoStop}%
\bibitem [{\citenamefont {den Nijs}(1988)}]{DenNijs}%
  \BibitemOpen
  \bibfield  {author} {\bibinfo {author} {\bibfnamefont {Marcel}\ \bibnamefont {den Nijs}},\ }\bibfield  {title} {\enquote {\bibinfo {title} {The domain wall theory of two-dimensional commensurate-incommensurate phase transitions},}\ }\href@noop {} {\bibfield  {journal} {\bibinfo  {journal} {Phase Transitions and Critical Phenomena}\ }\textbf {\bibinfo {volume} {12}},\ \bibinfo {pages} {219} (\bibinfo {year} {1988})}\BibitemShut {NoStop}%
\bibitem [{\citenamefont {Majumdar}\ and\ \citenamefont {Ghosh}(1969)}]{Majumdar1969}%
  \BibitemOpen
  \bibfield  {author} {\bibinfo {author} {\bibfnamefont {Chanchal~K.}\ \bibnamefont {Majumdar}}\ and\ \bibinfo {author} {\bibfnamefont {Dipan~K.}\ \bibnamefont {Ghosh}},\ }\bibfield  {title} {\enquote {\bibinfo {title} {{On Next‐Nearest‐Neighbor Interaction in Linear Chain. I}},}\ }\href {\doibase 10.1063/1.1664978} {\bibfield  {journal} {\bibinfo  {journal} {Journal of Mathematical Physics}\ }\textbf {\bibinfo {volume} {10}},\ \bibinfo {pages} {1388--1398} (\bibinfo {year} {1969})}\BibitemShut {NoStop}%
\bibitem [{\citenamefont {White}\ and\ \citenamefont {Affleck}(1996)}]{White1996}%
  \BibitemOpen
  \bibfield  {author} {\bibinfo {author} {\bibfnamefont {Steven~R.}\ \bibnamefont {White}}\ and\ \bibinfo {author} {\bibfnamefont {Ian}\ \bibnamefont {Affleck}},\ }\bibfield  {title} {\enquote {\bibinfo {title} {Dimerization and incommensurate spiral spin correlations in the zigzag spin chain: Analogies to the kondo lattice},}\ }\href {\doibase 10.1103/PhysRevB.54.9862} {\bibfield  {journal} {\bibinfo  {journal} {Phys. Rev. B}\ }\textbf {\bibinfo {volume} {54}},\ \bibinfo {pages} {9862--9869} (\bibinfo {year} {1996})}\BibitemShut {NoStop}%
\bibitem [{\citenamefont {Gozel}\ \emph {et~al.}(2019)\citenamefont {Gozel}, \citenamefont {Poilblanc}, \citenamefont {Affleck},\ and\ \citenamefont {Mila}}]{Gozel2019}%
  \BibitemOpen
  \bibfield  {author} {\bibinfo {author} {\bibfnamefont {Samuel}\ \bibnamefont {Gozel}}, \bibinfo {author} {\bibfnamefont {Didier}\ \bibnamefont {Poilblanc}}, \bibinfo {author} {\bibfnamefont {Ian}\ \bibnamefont {Affleck}}, \ and\ \bibinfo {author} {\bibfnamefont {Frédéric}\ \bibnamefont {Mila}},\ }\bibfield  {title} {\enquote {\bibinfo {title} {Novel families of su(n) aklt states with arbitrary self-conjugate edge states},}\ }\href {\doibase https://doi.org/10.1016/j.nuclphysb.2019.114663} {\bibfield  {journal} {\bibinfo  {journal} {Nuclear Physics B}\ }\textbf {\bibinfo {volume} {945}},\ \bibinfo {pages} {114663} (\bibinfo {year} {2019})}\BibitemShut {NoStop}%
\bibitem [{Note1()}]{Note1}%
  \BibitemOpen
  \bibinfo {note} {We can always choose $H_2 = \DOTSB \sum@ \slimits@ \limits _j h_{j+\protect \frac {1}{2}}^2$ if $h_{j+\protect \frac {1}{2}} $ is not positive definite.}\BibitemShut {Stop}%
\bibitem [{\citenamefont {Jutho}\ \emph {et~al.}(2024)\citenamefont {Jutho}, \citenamefont {Lukas}, \citenamefont {Hauru}, \citenamefont {maartenvd}, \citenamefont {ho~oto}, \citenamefont {Gertian}, \citenamefont {Burgelman}, \citenamefont {tangwei94}, \citenamefont {TagBot}, \citenamefont {Carlström}, \citenamefont {Xiaoyu},\ and\ \citenamefont {qmortier}}]{TensorKit}%
  \BibitemOpen
  \bibfield  {author} {\bibinfo {author} {\bibnamefont {Jutho}}, \bibinfo {author} {\bibnamefont {Lukas}}, \bibinfo {author} {\bibfnamefont {Markus}\ \bibnamefont {Hauru}}, \bibinfo {author} {\bibnamefont {maartenvd}}, \bibinfo {author} {\bibnamefont {ho~oto}}, \bibinfo {author} {\bibnamefont {Gertian}}, \bibinfo {author} {\bibfnamefont {Lander}\ \bibnamefont {Burgelman}}, \bibinfo {author} {\bibnamefont {tangwei94}}, \bibinfo {author} {\bibfnamefont {Julia}\ \bibnamefont {TagBot}}, \bibinfo {author} {\bibfnamefont {Stefanos}\ \bibnamefont {Carlström}}, \bibinfo {author} {\bibnamefont {Xiaoyu}}, \ and\ \bibinfo {author} {\bibnamefont {qmortier}},\ }\href {\doibase 10.5281/zenodo.12633850} {\enquote {\bibinfo {title} {Jutho/tensorkit.jl: v0.12.5},}\ } (\bibinfo {year} {2024})\BibitemShut {NoStop}%
\bibitem [{\citenamefont {Van~Damme}\ \emph {et~al.}(2024)\citenamefont {Van~Damme}, \citenamefont {Devos},\ and\ \citenamefont {Haegeman}}]{MPSKIT}%
  \BibitemOpen
  \bibfield  {author} {\bibinfo {author} {\bibfnamefont {Maarten}\ \bibnamefont {Van~Damme}}, \bibinfo {author} {\bibfnamefont {Lukas}\ \bibnamefont {Devos}}, \ and\ \bibinfo {author} {\bibfnamefont {Jutho}\ \bibnamefont {Haegeman}},\ }\href {\doibase 10.5281/zenodo.13175814} {\enquote {\bibinfo {title} {Mpskit},}\ } (\bibinfo {year} {2024})\BibitemShut {NoStop}%
\bibitem [{\citenamefont {Alex}\ \emph {et~al.}(2011)\citenamefont {Alex}, \citenamefont {Kalus}, \citenamefont {Huckleberry},\ and\ \citenamefont {von Delft}}]{Alex2011}%
  \BibitemOpen
  \bibfield  {author} {\bibinfo {author} {\bibfnamefont {Arne}\ \bibnamefont {Alex}}, \bibinfo {author} {\bibfnamefont {Matthias}\ \bibnamefont {Kalus}}, \bibinfo {author} {\bibfnamefont {Alan}\ \bibnamefont {Huckleberry}}, \ and\ \bibinfo {author} {\bibfnamefont {Jan}\ \bibnamefont {von Delft}},\ }\bibfield  {title} {\enquote {\bibinfo {title} {{A numerical algorithm for the explicit calculation of SU(N) and SL(N, C) Clebsch–Gordan coefficients}},}\ }\href {\doibase 10.1063/1.3521562} {\bibfield  {journal} {\bibinfo  {journal} {Journal of Mathematical Physics}\ }\textbf {\bibinfo {volume} {52}},\ \bibinfo {pages} {023507} (\bibinfo {year} {2011})}\BibitemShut {NoStop}%
\bibitem [{\citenamefont {White}(1992)}]{WhiteDMRG}%
  \BibitemOpen
  \bibfield  {author} {\bibinfo {author} {\bibfnamefont {Steven~R.}\ \bibnamefont {White}},\ }\bibfield  {title} {\enquote {\bibinfo {title} {Density matrix formulation for quantum renormalization groups},}\ }\href {\doibase 10.1103/PhysRevLett.69.2863} {\bibfield  {journal} {\bibinfo  {journal} {Phys. Rev. Lett.}\ }\textbf {\bibinfo {volume} {69}},\ \bibinfo {pages} {2863--2866} (\bibinfo {year} {1992})}\BibitemShut {NoStop}%
\bibitem [{\citenamefont {Vidal}(2007)}]{VidaliDMRG}%
  \BibitemOpen
  \bibfield  {author} {\bibinfo {author} {\bibfnamefont {G.}~\bibnamefont {Vidal}},\ }\bibfield  {title} {\enquote {\bibinfo {title} {Classical simulation of infinite-size quantum lattice systems in one spatial dimension},}\ }\href {\doibase 10.1103/PhysRevLett.98.070201} {\bibfield  {journal} {\bibinfo  {journal} {Phys. Rev. Lett.}\ }\textbf {\bibinfo {volume} {98}},\ \bibinfo {pages} {070201} (\bibinfo {year} {2007})}\BibitemShut {NoStop}%
\bibitem [{\citenamefont {McCulloch}()}]{McCullochiDMRG}%
  \BibitemOpen
  \bibfield  {author} {\bibinfo {author} {\bibfnamefont {I.~P.}\ \bibnamefont {McCulloch}},\ }\bibfield  {title} {\enquote {\bibinfo {title} {Infinite size density matrix renormalization group, revisited},}\ }\href {https://api.semanticscholar.org/CorpusID:118257570} {\ }\Eprint {http://arxiv.org/abs/0804.2509} {arXiv:0804.2509} \BibitemShut {NoStop}%
\bibitem [{\citenamefont {Zauner-Stauber}\ \emph {et~al.}(2018)\citenamefont {Zauner-Stauber}, \citenamefont {Vanderstraeten}, \citenamefont {Fishman}, \citenamefont {Verstraete},\ and\ \citenamefont {Haegeman}}]{VUMPS}%
  \BibitemOpen
  \bibfield  {author} {\bibinfo {author} {\bibfnamefont {V.}~\bibnamefont {Zauner-Stauber}}, \bibinfo {author} {\bibfnamefont {L.}~\bibnamefont {Vanderstraeten}}, \bibinfo {author} {\bibfnamefont {M.~T.}\ \bibnamefont {Fishman}}, \bibinfo {author} {\bibfnamefont {F.}~\bibnamefont {Verstraete}}, \ and\ \bibinfo {author} {\bibfnamefont {J.}~\bibnamefont {Haegeman}},\ }\bibfield  {title} {\enquote {\bibinfo {title} {Variational optimization algorithms for uniform matrix product states},}\ }\href {\doibase 10.1103/PhysRevB.97.045145} {\bibfield  {journal} {\bibinfo  {journal} {Phys. Rev. B}\ }\textbf {\bibinfo {volume} {97}},\ \bibinfo {pages} {045145} (\bibinfo {year} {2018})}\BibitemShut {NoStop}%
\bibitem [{\citenamefont {Hauru}\ \emph {et~al.}(2021)\citenamefont {Hauru}, \citenamefont {Damme},\ and\ \citenamefont {Haegeman}}]{GrassmannDescent}%
  \BibitemOpen
  \bibfield  {author} {\bibinfo {author} {\bibfnamefont {Markus}\ \bibnamefont {Hauru}}, \bibinfo {author} {\bibfnamefont {Maarten~Van}\ \bibnamefont {Damme}}, \ and\ \bibinfo {author} {\bibfnamefont {Jutho}\ \bibnamefont {Haegeman}},\ }\bibfield  {title} {\enquote {\bibinfo {title} {{Riemannian optimization of isometric tensor networks}},}\ }\href {\doibase 10.21468/SciPostPhys.10.2.040} {\bibfield  {journal} {\bibinfo  {journal} {SciPost Phys.}\ }\textbf {\bibinfo {volume} {10}},\ \bibinfo {pages} {040} (\bibinfo {year} {2021})}\BibitemShut {NoStop}%
\bibitem [{\citenamefont {Hubig}\ \emph {et~al.}(2015)\citenamefont {Hubig}, \citenamefont {McCulloch}, \citenamefont {Schollw\"ock},\ and\ \citenamefont {Wolf}}]{SubspaceExpansion}%
  \BibitemOpen
  \bibfield  {author} {\bibinfo {author} {\bibfnamefont {C.}~\bibnamefont {Hubig}}, \bibinfo {author} {\bibfnamefont {I.~P.}\ \bibnamefont {McCulloch}}, \bibinfo {author} {\bibfnamefont {U.}~\bibnamefont {Schollw\"ock}}, \ and\ \bibinfo {author} {\bibfnamefont {F.~A.}\ \bibnamefont {Wolf}},\ }\bibfield  {title} {\enquote {\bibinfo {title} {Strictly single-site dmrg algorithm with subspace expansion},}\ }\href {\doibase 10.1103/PhysRevB.91.155115} {\bibfield  {journal} {\bibinfo  {journal} {Phys. Rev. B}\ }\textbf {\bibinfo {volume} {91}},\ \bibinfo {pages} {155115} (\bibinfo {year} {2015})}\BibitemShut {NoStop}%
\bibitem [{\citenamefont {Ornstein}\ and\ \citenamefont {Zernike}(1914)}]{Ornstein1914}%
  \BibitemOpen
  \bibfield  {author} {\bibinfo {author} {\bibfnamefont {L.~S.}\ \bibnamefont {Ornstein}}\ and\ \bibinfo {author} {\bibfnamefont {F.}~\bibnamefont {Zernike}},\ }\bibfield  {title} {\enquote {\bibinfo {title} {Accidental deviations of density and opalescence at the critical point of a single substance},}\ }\href@noop {} {\bibfield  {journal} {\bibinfo  {journal} {Proc. Acad. Sci. Amsterdam}\ }\textbf {\bibinfo {volume} {17}} (\bibinfo {year} {1914})}\BibitemShut {NoStop}%
\bibitem [{\citenamefont {Zauner}\ \emph {et~al.}(2015)\citenamefont {Zauner}, \citenamefont {Draxler}, \citenamefont {Vanderstraeten}, \citenamefont {Degroote}, \citenamefont {Haegeman}, \citenamefont {Rams}, \citenamefont {Stojevic}, \citenamefont {Schuch},\ and\ \citenamefont {Verstraete}}]{Zauner2015}%
  \BibitemOpen
  \bibfield  {author} {\bibinfo {author} {\bibfnamefont {V}~\bibnamefont {Zauner}}, \bibinfo {author} {\bibfnamefont {D}~\bibnamefont {Draxler}}, \bibinfo {author} {\bibfnamefont {L}~\bibnamefont {Vanderstraeten}}, \bibinfo {author} {\bibfnamefont {M}~\bibnamefont {Degroote}}, \bibinfo {author} {\bibfnamefont {J}~\bibnamefont {Haegeman}}, \bibinfo {author} {\bibfnamefont {M~M}\ \bibnamefont {Rams}}, \bibinfo {author} {\bibfnamefont {V}~\bibnamefont {Stojevic}}, \bibinfo {author} {\bibfnamefont {N}~\bibnamefont {Schuch}}, \ and\ \bibinfo {author} {\bibfnamefont {F}~\bibnamefont {Verstraete}},\ }\bibfield  {title} {\enquote {\bibinfo {title} {Transfer matrices and excitations with matrix product states},}\ }\href {\doibase 10.1088/1367-2630/17/5/053002} {\bibfield  {journal} {\bibinfo  {journal} {New Journal of Physics}\ }\textbf {\bibinfo {volume} {17}},\ \bibinfo {pages} {053002} (\bibinfo {year} {2015})}\BibitemShut {NoStop}%
\bibitem [{\citenamefont {Rams}\ \emph {et~al.}(2018)\citenamefont {Rams}, \citenamefont {Czarnik},\ and\ \citenamefont {Cincio}}]{Rams2018}%
  \BibitemOpen
  \bibfield  {author} {\bibinfo {author} {\bibfnamefont {Marek~M.}\ \bibnamefont {Rams}}, \bibinfo {author} {\bibfnamefont {Piotr}\ \bibnamefont {Czarnik}}, \ and\ \bibinfo {author} {\bibfnamefont {Lukasz}\ \bibnamefont {Cincio}},\ }\bibfield  {title} {\enquote {\bibinfo {title} {Precise extrapolation of the correlation function asymptotics in uniform tensor network states with application to the bose-hubbard and xxz models},}\ }\href {\doibase 10.1103/PhysRevX.8.041033} {\bibfield  {journal} {\bibinfo  {journal} {Phys. Rev. X}\ }\textbf {\bibinfo {volume} {8}},\ \bibinfo {pages} {041033} (\bibinfo {year} {2018})}\BibitemShut {NoStop}%
\bibitem [{\citenamefont {Vanhecke}\ \emph {et~al.}(2019)\citenamefont {Vanhecke}, \citenamefont {Haegeman}, \citenamefont {Van~Acoleyen}, \citenamefont {Vanderstraeten},\ and\ \citenamefont {Verstraete}}]{Vanhecke2019}%
  \BibitemOpen
  \bibfield  {author} {\bibinfo {author} {\bibfnamefont {Bram}\ \bibnamefont {Vanhecke}}, \bibinfo {author} {\bibfnamefont {Jutho}\ \bibnamefont {Haegeman}}, \bibinfo {author} {\bibfnamefont {Karel}\ \bibnamefont {Van~Acoleyen}}, \bibinfo {author} {\bibfnamefont {Laurens}\ \bibnamefont {Vanderstraeten}}, \ and\ \bibinfo {author} {\bibfnamefont {Frank}\ \bibnamefont {Verstraete}},\ }\bibfield  {title} {\enquote {\bibinfo {title} {Scaling hypothesis for matrix product states},}\ }\href {\doibase 10.1103/PhysRevLett.123.250604} {\bibfield  {journal} {\bibinfo  {journal} {Phys. Rev. Lett.}\ }\textbf {\bibinfo {volume} {123}},\ \bibinfo {pages} {250604} (\bibinfo {year} {2019})}\BibitemShut {NoStop}%
\bibitem [{\citenamefont {Greiter}\ and\ \citenamefont {Rachel}(2007)}]{Greiter2007}%
  \BibitemOpen
  \bibfield  {author} {\bibinfo {author} {\bibfnamefont {Martin}\ \bibnamefont {Greiter}}\ and\ \bibinfo {author} {\bibfnamefont {Stephan}\ \bibnamefont {Rachel}},\ }\bibfield  {title} {\enquote {\bibinfo {title} {Valence bond solids for $\mathrm{SU}(n)$ spin chains: Exact models, spinon confinement, and the haldane gap},}\ }\href {\doibase 10.1103/PhysRevB.75.184441} {\bibfield  {journal} {\bibinfo  {journal} {Phys. Rev. B}\ }\textbf {\bibinfo {volume} {75}},\ \bibinfo {pages} {184441} (\bibinfo {year} {2007})}\BibitemShut {NoStop}%
\bibitem [{\citenamefont {Greiter}\ \emph {et~al.}(2007)\citenamefont {Greiter}, \citenamefont {Rachel},\ and\ \citenamefont {Schuricht}}]{Greiter2007b}%
  \BibitemOpen
  \bibfield  {author} {\bibinfo {author} {\bibfnamefont {Martin}\ \bibnamefont {Greiter}}, \bibinfo {author} {\bibfnamefont {Stephan}\ \bibnamefont {Rachel}}, \ and\ \bibinfo {author} {\bibfnamefont {Dirk}\ \bibnamefont {Schuricht}},\ }\bibfield  {title} {\enquote {\bibinfo {title} {Exact results for su(3) spin chains: Trimer states, valence bond solids, and their parent hamiltonians},}\ }\href {\doibase 10.1103/PhysRevB.75.060401} {\bibfield  {journal} {\bibinfo  {journal} {Phys. Rev. B}\ }\textbf {\bibinfo {volume} {75}},\ \bibinfo {pages} {060401} (\bibinfo {year} {2007})}\BibitemShut {NoStop}%
\bibitem [{\citenamefont {Katsura}\ \emph {et~al.}(2008)\citenamefont {Katsura}, \citenamefont {Hirano},\ and\ \citenamefont {Korepin}}]{Katsura2008}%
  \BibitemOpen
  \bibfield  {author} {\bibinfo {author} {\bibfnamefont {Hosho}\ \bibnamefont {Katsura}}, \bibinfo {author} {\bibfnamefont {Takaaki}\ \bibnamefont {Hirano}}, \ and\ \bibinfo {author} {\bibfnamefont {Vladimir~E}\ \bibnamefont {Korepin}},\ }\bibfield  {title} {\enquote {\bibinfo {title} {Entanglement in an su(n) valence-bond-solid state},}\ }\href {\doibase 10.1088/1751-8113/41/13/135304} {\bibfield  {journal} {\bibinfo  {journal} {Journal of Physics A: Mathematical and Theoretical}\ }\textbf {\bibinfo {volume} {41}},\ \bibinfo {pages} {135304} (\bibinfo {year} {2008})}\BibitemShut {NoStop}%
\bibitem [{\citenamefont {Morimoto}\ \emph {et~al.}(2014)\citenamefont {Morimoto}, \citenamefont {Ueda}, \citenamefont {Momoi},\ and\ \citenamefont {Furusaki}}]{Morimoto2014}%
  \BibitemOpen
  \bibfield  {author} {\bibinfo {author} {\bibfnamefont {Takahiro}\ \bibnamefont {Morimoto}}, \bibinfo {author} {\bibfnamefont {Hiroshi}\ \bibnamefont {Ueda}}, \bibinfo {author} {\bibfnamefont {Tsutomu}\ \bibnamefont {Momoi}}, \ and\ \bibinfo {author} {\bibfnamefont {Akira}\ \bibnamefont {Furusaki}},\ }\bibfield  {title} {\enquote {\bibinfo {title} {${\mathbb{z}}_{3}$ symmetry-protected topological phases in the su(3) aklt model},}\ }\href {\doibase 10.1103/PhysRevB.90.235111} {\bibfield  {journal} {\bibinfo  {journal} {Phys. Rev. B}\ }\textbf {\bibinfo {volume} {90}},\ \bibinfo {pages} {235111} (\bibinfo {year} {2014})}\BibitemShut {NoStop}%
\bibitem [{\citenamefont {Gozel}\ \emph {et~al.}(2020)\citenamefont {Gozel}, \citenamefont {Nataf},\ and\ \citenamefont {Mila}}]{Gozel2020}%
  \BibitemOpen
  \bibfield  {author} {\bibinfo {author} {\bibfnamefont {Samuel}\ \bibnamefont {Gozel}}, \bibinfo {author} {\bibfnamefont {Pierre}\ \bibnamefont {Nataf}}, \ and\ \bibinfo {author} {\bibfnamefont {Fr\'ed\'eric}\ \bibnamefont {Mila}},\ }\bibfield  {title} {\enquote {\bibinfo {title} {Haldane gap of the three-box symmetric su(3) chain},}\ }\href {\doibase 10.1103/PhysRevLett.125.057202} {\bibfield  {journal} {\bibinfo  {journal} {Phys. Rev. Lett.}\ }\textbf {\bibinfo {volume} {125}},\ \bibinfo {pages} {057202} (\bibinfo {year} {2020})}\BibitemShut {NoStop}%
\bibitem [{\citenamefont {Devos}\ \emph {et~al.}(2022)\citenamefont {Devos}, \citenamefont {Vanderstraeten},\ and\ \citenamefont {Verstraete}}]{Devos2022}%
  \BibitemOpen
  \bibfield  {author} {\bibinfo {author} {\bibfnamefont {Lukas}\ \bibnamefont {Devos}}, \bibinfo {author} {\bibfnamefont {Laurens}\ \bibnamefont {Vanderstraeten}}, \ and\ \bibinfo {author} {\bibfnamefont {Frank}\ \bibnamefont {Verstraete}},\ }\bibfield  {title} {\enquote {\bibinfo {title} {Haldane gap in the su(3) [3 0 0] heisenberg chain},}\ }\href {\doibase 10.1103/PhysRevB.106.155103} {\bibfield  {journal} {\bibinfo  {journal} {Phys. Rev. B}\ }\textbf {\bibinfo {volume} {106}},\ \bibinfo {pages} {155103} (\bibinfo {year} {2022})}\BibitemShut {NoStop}%
\bibitem [{\citenamefont {Lajkó}\ \emph {et~al.}(2017)\citenamefont {Lajkó}, \citenamefont {Wamer}, \citenamefont {Mila},\ and\ \citenamefont {Affleck}}]{Lajko2017}%
  \BibitemOpen
  \bibfield  {author} {\bibinfo {author} {\bibfnamefont {Miklós}\ \bibnamefont {Lajkó}}, \bibinfo {author} {\bibfnamefont {Kyle}\ \bibnamefont {Wamer}}, \bibinfo {author} {\bibfnamefont {Frédéric}\ \bibnamefont {Mila}}, \ and\ \bibinfo {author} {\bibfnamefont {Ian}\ \bibnamefont {Affleck}},\ }\bibfield  {title} {\enquote {\bibinfo {title} {Generalization of the haldane conjecture to su(3) chains},}\ }\href {\doibase https://doi.org/10.1016/j.nuclphysb.2017.09.015} {\bibfield  {journal} {\bibinfo  {journal} {Nuclear Physics B}\ }\textbf {\bibinfo {volume} {924}},\ \bibinfo {pages} {508--577} (\bibinfo {year} {2017})}\BibitemShut {NoStop}%
\bibitem [{\citenamefont {Wamer}\ \emph {et~al.}(2020)\citenamefont {Wamer}, \citenamefont {Lajkó}, \citenamefont {Mila},\ and\ \citenamefont {Affleck}}]{Wamer2020}%
  \BibitemOpen
  \bibfield  {author} {\bibinfo {author} {\bibfnamefont {Kyle}\ \bibnamefont {Wamer}}, \bibinfo {author} {\bibfnamefont {Miklós}\ \bibnamefont {Lajkó}}, \bibinfo {author} {\bibfnamefont {Frédéric}\ \bibnamefont {Mila}}, \ and\ \bibinfo {author} {\bibfnamefont {Ian}\ \bibnamefont {Affleck}},\ }\bibfield  {title} {\enquote {\bibinfo {title} {Generalization of the haldane conjecture to su(n) chains},}\ }\href {\doibase https://doi.org/10.1016/j.nuclphysb.2020.114932} {\bibfield  {journal} {\bibinfo  {journal} {Nuclear Physics B}\ }\textbf {\bibinfo {volume} {952}},\ \bibinfo {pages} {114932} (\bibinfo {year} {2020})}\BibitemShut {NoStop}%
\bibitem [{\citenamefont {Nataf}\ and\ \citenamefont {Mila}(2018)}]{Nataf2018}%
  \BibitemOpen
  \bibfield  {author} {\bibinfo {author} {\bibfnamefont {Pierre}\ \bibnamefont {Nataf}}\ and\ \bibinfo {author} {\bibfnamefont {Fr\'ed\'eric}\ \bibnamefont {Mila}},\ }\bibfield  {title} {\enquote {\bibinfo {title} {Density matrix renormalization group simulations of su($n$) heisenberg chains using standard young tableaus: Fundamental representation and comparison with a finite-size bethe ansatz},}\ }\href {\doibase 10.1103/PhysRevB.97.134420} {\bibfield  {journal} {\bibinfo  {journal} {Phys. Rev. B}\ }\textbf {\bibinfo {volume} {97}},\ \bibinfo {pages} {134420} (\bibinfo {year} {2018})}\BibitemShut {NoStop}%
\end{thebibliography}%
